\documentclass[aps,pre,twocolumn,floatfix,letterpaper,amssymb,superscriptaddress]{revtex4-1}
\usepackage{graphicx}
\usepackage[intlimits]{amsmath}

\makeatletter
\def\dleft#1#2\dright{\begingroup%
\def\ts@r{\nulldelimiterspace=0pt \mathsurround=0pt}%
\let\@hat=#1%
\def\sht@im{#2}%
\def\@t{{\mathchoice{\def\@fen{\displaystyle}\k@fel}%
{\def\@fen{\textstyle}\k@fel}%
{\def\@fen{\scriptstyle}\k@fel}%
{\def\@fen{\scriptscriptstyle}\k@fel}}}%
\def\g@rin{\ts@r\left\@hat\vphantom{\sht@im}\right.}%
\def\k@fel{\setbox0=\hbox{$\@fen\g@rin$}\hbox{%
$\@fen \kern.3875\wd0 \copy0 \kern-.3875\wd0%
\llap{\copy0}\kern.3875\wd0$}}%
\def\pt@h{\mathopen\@t}\pt@h\sht@im%
\dright}%
\def\dright#1{\let\@hat=#1%
\def\st@m{\mathclose\@t}%
\st@m\endgroup}
\makeatother

\begin{document}

\title{Inelastic Microwave Photon Scattering off a Quantum Impurity in a Josephson-Junction Array}

\author{Moshe Goldstein}
\affiliation{Department of Physics, Yale University, New Haven, CT 06520, USA}

\author{Michel H. Devoret}
\affiliation{Department of Physics, Yale University, New Haven, CT 06520, USA}
\affiliation{Departments of Applied Physics, Yale University, New Haven, CT 06520, USA}

\author{Manuel Houzet}
\affiliation{SPSMS, UMR-E 9001, CEA-INAC/UJF-Grenoble 1, F-38054 Grenoble, France}

\author{Leonid I. Glazman}
\affiliation{Department of Physics, Yale University, New Haven, CT 06520, USA}
\affiliation{Departments of Applied Physics, Yale University, New Haven, CT 06520, USA}

\begin{abstract}
Quantum fluctuations in an anharmonic superconducting circuit enable frequency conversion of individual incoming  photons. This effect, linear in the photon beam intensity, leads to ramifications for the standard input-output circuit theory.
We consider an extreme case of anharmonicity in which photons scatter off
a small set of weak links within a Josephson junction array. We show that this quantum impurity displays Kondo physics and evaluate the elastic and inelastic photon scattering cross sections. These cross sections reveal many-body properties of the Kondo problem that are hard to access in its traditional fermionic version.
\end{abstract}


\maketitle

Propagation of small-amplitude electromagnetic waves through an optical system or a passive microwave circuit is conventionally described in terms of transmission and reflection amplitudes, or, equivalently, complex admittances. Considered classically, the wave propagation can be calculated using input-output theory \cite{ujihara,clerk10}.
In the absence of dissipation, the transmission $t(\omega)$ and reflection $r(\omega)$ amplitudes for a photon of frequency $\omega$ satisfy the unitarity condition, $|t(\omega)|^2+|r(\omega)|^2=1$.
It is often tacitly assumed that this description applies in the quantum limit too.
While this is indeed true if the circuit is harmonic,
the presence of anharmonic elements 
modifies the picture qualitatively:
a photon of energy $\hbar\omega$ may ``split'' into several ones of smaller energy; unitarity is violated in the elastic channel, $|t(\omega)|^2+|r(\omega)|^2<1$. The photon frequency conversion results in a finite dissipative part of the admittances
despite the system being free of dissipative elements. These features appear in a quantum circuit containing even a single or a small group of anharmonic elements, a ``quantum impurity''.

\begin{figure}[b]
\includegraphics[width=8.5cm,height=!]{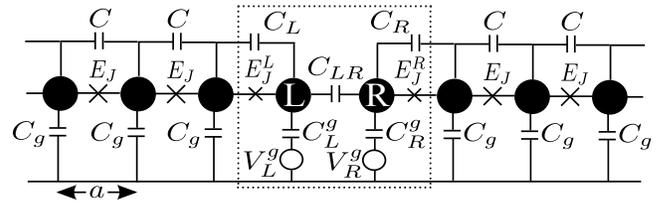}
\caption{\label{fig:system}
Diagram of the system. The dotted box surrounds the quantum impurity.
See the text for further details.
}
\end{figure}

In this paper we consider the propagation of microwave photons (oscillations of charge and superconducting phase) along an array of Josephson junctions interrupted by a capacitive element; see Fig.~\ref{fig:system}.
If Josephson energies were all large with respect to charging energies for each of the tunnel junctions, the system would be effectively harmonic,
and photon scattering off the central capacitive link would be purely elastic.
We will rather assume the Josephson energy to be large for all the junctions \emph{except} for the two closest to the capacitive link. These two junctions, together with the two superconducting islands they single out, form a quantum impurity which causes inelastic photon scattering.
The quantum impurity is of the Kondo variety
\cite{leclair97,camalet04,garciaripoll08,lehur12,hewson}, where the two values of the polarization charge of the said two islands play the role of the Kondo spin.
However, photon scattering is quite different from electron scattering in the conventional Kondo problem \cite{garst05}. We find that the photon elastic transmission and reflection coefficients, as well as the total inelastic scattering cross section $\gamma (\omega)$, are related to the local ``spin'' susceptibility $\chi_{zz}(\omega)$.
We then study the spectrum $\gamma (\omega^\prime | \omega)$ of photons at frequency $\omega^\prime$ generated by inelastic processes from incoming photons at frequency $\omega$.
The spectrum peaks as a function of $\omega^\prime$ at the Kondo energy scale. At $\omega-\omega^\prime \ll T_K$ or $\omega^\prime \ll T_K$ the behavior of $\gamma (\omega^\prime |\omega)$ provides direct access to corrections to the Nozi\`{e}res fixed-point Hamiltonian.
We provide technical details in the Supplemental Material (SM) \cite{sm}.

Assuming that the superconducting gap is larger than any other energy scale, the only relevant degrees of freedom are the number of Cooper pairs $n_i$ on island $i$ and the corresponding superconducting phase $\varphi_i$, obeying $[\varphi_i, n_j]=i \delta_{ij}$. The array Hamiltonian is
\begin{equation} \label{eqn:h_array}
  H = \sum_{i,j}
  \left[
  2e^2 \left(n_i - n_i^{0}\right)
  \left(\mathsf{C}^{-1}\right)_{ij}
  \left(n_j - n_j^{0}\right)
  -\mathsf{E}_J^{ij} \cos ( \varphi_i - \varphi_j )
  \right],
\end{equation}
where $\mathsf{E}_J^{ij}$ and $\mathsf{C}_{ij}$ are the matrices of Josephson couplings and capacitances, respectively. We will assume nearest-neighbor Josephson couplings, and ground- and nearest-neighbor capacitances, whose values can be inferred from Fig.~\ref{fig:system}. The gate-induced charge offset on the $i$th island is $n_i^{0} = C^g_i V^g_i/(2 e)$ with $V_i^g$ and $C^g_i$ being the gate voltage and capacitance to the ground, respectively.

Away from the quantum impurity the array is uniform: except for the quantum impurity islands, all Josephson couplings are $E_J$, and all capacitances to the ground and junction capacitances are $C_g$ and  $C$, respectively.
Properties of the uniform array are controlled by two ratios, $E_J/E_{C_g}$ and $E_J/E_C$, of $E_J$ and two charging energies, $E_C=(2e)^2/(2C)$ and $E_{C_g}=(2e)^2/(2C_g)$. Typically $C/C_g\gg 1$ (it is $\sim 10^2$ in~\cite{manucharyan09}). That allows one to have the impedance of the array $Z=[\hbar/(2e)^2]\sqrt{2 E_{C_g}/E_J}$ on the order of the resistance quantum $R_Q=\pi\hbar/(2e^2)$, while keeping the amplitude of phase slips $\mathcal{A} \sim e^{-\sqrt{32E_J/E_C}}$ exponentially small \cite{manucharyan09}.
In an array of length $L \lesssim a/\mathcal{A}$ ($a$ is the array spacing) the Josephson energy can thus be replaced by a quadratic term.
In addition, in the long wavelength limit we may use a continuum description for the array \cite{gogolin} (except for the impurity) in terms of Bose fields $\phi_\ell(x)$  and $\rho_\ell(x)$ which represent, respectively, the superconducting phase (whose gradient is proportional to the electric current) and charge density (in units of $- 2 e$ per period of the array) in lead $\ell=L,R$, obeying
$[ \phi_\ell(x), \rho_{\ell^\prime}(x^\prime) ]  =
i \delta_{\ell \ell^\prime} \delta(x-x^\prime)$,
\begin{equation} \label{eqn:h_leads}
  H_\text{leads} = \negthickspace \negthickspace
  \sum_{\ell=L,R} \negthickspace
  \frac{v}{2\pi} \int_0^\infty \negthickspace \left\{ g \left[ \partial_x \phi_\ell(x) \right]^2 + \frac{1}{g} \left[ \pi \rho_\ell(x) \right]^2 \right\} \text{d}x.
\end{equation}
The array is characterized by the velocity of plasmons $v = a \sqrt{2 E_J E_{C_g}}$, and by $g = R_Q/ (2 Z)$.
$C$ does not affect excitations of wavelengths well exceeding $a \sqrt{C/C_g}$. Thus, the linear dispersion waveguide Hamiltonian (\ref{eqn:h_leads}) is limited to frequencies within a bandwidth $\omega_0 \sim (v/a) \sqrt{C_g/C}$
(See SM, Sec.~SM.A \cite{sm}).

Let us now turn to the quantum impurity, islands $L$ and $R$ in the dotted box in Fig.~\ref{fig:system}.
We derive its low-energy Hamiltonian under the realistic assumptions $C_{LR} \sim C \gg C^g_L,C^g_R \sim C_g$ and $C_L, C_R \sim \sqrt{C C_g}$
(See SM, Sec.~SM.A \cite{sm}).
When the charging energy $E_C^\text{imp}=(2e)^2/[2 (\tilde{C}_L+\tilde{C}_R)]$,
with $1/\tilde{C}_\ell = 1/C_\ell + 1/\sqrt{C C_g}$,
is large with respect to the Josephson energies $E_J^{L,R}$, the total impurity charge $n_L+n_R$ is quantized.
If the gate voltages are set to $(C^g_L V^g_L + C^g_R V^g_R)/(2e)=1$,
then to lowest order in $E_J^{L,R}$ the islands are restricted to the two charging states
$| 0_L,1_R \rangle$ and 
$| 1_L,0_R \rangle$. 
We label these two configurations by the states of a pseudospin, $S_z = (n_L-n_R)/2 = \pm 1/2$, so that
$S_+ = | 1_L,0_R \rangle \langle 0_L,1_R |$ and $S_-=(S_+)^\dagger$. 
Finite $E_J^{L,R}$ enables switching between these two states through virtual states with energies of order $E_C^\text{imp}$.
Eliminating these by a Schrieffer-Wolff transformation leads to an effective low-energy Hamiltonian
(See SM, Sec.~SM.A \cite{sm}),
\begin{align}
\label{eqn:h_lm}
H_\text{imp}
= &
- \frac{E_J^{LR}}{2} \left\{ e^{-i[\phi_L(0)-\phi_R(0)]} S_+
+ e^{i[\phi_L(0)-\phi_R(0)]} S_- \right\}
\nonumber \\
&
+\frac{(2e)^2}{C_g} \lambda_{LR} a \left[ \rho_L(0) - \rho_R(0) \right] S_z
- B_z S_z.
\end{align}
Here
\begin{equation}
\label{eqn:bz_ejlr}
E_J^{LR}= \frac{E_J^L E_J^R}{E_C^\text{imp}}, \quad
\frac{B_z}{2e} = \left( \frac{1}{2 C_{LR}} - \frac{\lambda_{LR}^2}{C_g} \right)
\left( C^g_L  V^g_L - C^g_R  V^g_R \right)
\end{equation}
and
$\lambda_{LR} = C_L C_R / [(C_L + C_R ) C_{LR}] \sim \sqrt{C_g/C} \ll 1$.
The first term in Eq.~(\ref{eqn:h_lm}) accounts for
flips of the pseudospin, which are accompanied by transfers of discrete charge $\pm 2e$ between the two leads \cite{fn:ejlr_renormalization}. 
The second term is a capacitive coupling between the impurity and the leads. The third represents the effect of a gate voltage bias between the impurity islands.
Hamiltonian~(\ref{eqn:h_lm}) clearly introduces anharmonicity into the system.

Applying the transformation $H \rightarrow \mathcal{U}^\dagger H \mathcal{U}$ with $\mathcal{U}= e^{-i[\phi_L(0)-\phi_R(0)] S_z}$, the Hamiltonian acquires the form of the spin-boson model with Ohmic dissipation \cite{leggett87,weiss}:
\begin{multline}
\label{eqn:h_sb}
  H_{SB} =
  \sum_{\lambda=c,s} \frac{v}{2\pi} \int_0^\infty \left\{ \left[ \partial_x \tilde{\phi}_\lambda(x) \right]^2 + \left[ \pi \tilde{\rho}_\lambda(x) \right]^2 \right\} \text{d}x
  \\
  - B_z S_z
  - E_J^{LR} S_x 
  - \pi v \alpha \tilde{\rho}_s(0) S_z,
\end{multline}
where $\tilde{\rho}_s(x) = [\alpha_L \rho_L(x) -\alpha_R \rho_R(x)]/(\alpha \sqrt{g})$ and $\tilde{\phi}_s(x) = \sqrt{g} [\alpha_L \phi_L(x) -\alpha_R \phi_R(x)]/\alpha$ are, respectively, the ``spin density'' and its canonically conjugate momentum field.
The ``charge density'' and its conjugate field,  $\tilde{\rho}_c(x) = [\alpha_R \rho_L(x) +\alpha_L \rho_R(x)]/(\alpha \sqrt{g})$ and $\tilde{\phi}_c(x) = \sqrt{g} [\alpha_R \phi_L(x) +\alpha_L \phi_R(x)]/\alpha$, decouple from the impurity spin.
The parameters $\alpha_{L,R}$ and the coupling parameter $\alpha$ in Eq.~(\ref{eqn:h_sb}) are given by  \cite{fn:alpha_lr}
\begin{equation}
\label{eqn:alpha}
  \alpha_{L} = \alpha_{R} = \frac{1}{\sqrt{g}} \left( 1 - \lambda_{LR} \right),
  \qquad 
  \alpha^2 = \alpha_L^2 + \alpha_R^2.
\end{equation}

The spin-boson Hamiltonian (\ref{eqn:h_sb}) is equivalent \cite{leggett87,weiss} to the single-channel Kondo model \cite{hewson}, describing a localized spin exchange-coupled to a bath of noninteracting spin-$1/2$ fermions with bandwidth $\omega_0$,
\begin{multline}
\label{eqn:h_k}
  H_K =
  \sum_{k,\sigma=\uparrow,\downarrow} v k c^\dagger_{k,\sigma} c_{k,\sigma}
  +\frac{I_z}{2L} S_z \sum_{k,\sigma,k^\prime,\sigma^\prime} c^\dagger_{k,\sigma} \tau^{z}_{\sigma,\sigma^\prime} c_{k^\prime,\sigma^\prime}
  \\
  + \frac{I_{xy}}{4L} S_{-} \sum_{k,\sigma,k^\prime,\sigma^\prime} c^\dagger_{k,\sigma} \tau^{+}_{\sigma,\sigma^\prime} c_{k^\prime,\sigma^\prime} + \text{H.c.}
  - B_z S_z,
\end{multline}
where $\tau^{i}_{\sigma,\sigma^\prime}$ 
are the Pauli matrices, $I_z = 2 \pi v (1 - \alpha/\sqrt{2})$,
and $I_{xy} = 2 \pi a E_J^{LR}$.
Given the smallness of $E_J^{LR}$ [cf.\ Eq.~(\ref{eqn:bz_ejlr})], isotropic exchange ($I_{xy}=I_z$) corresponds to $\alpha^2 \approx 2$ (i.e., $g \approx 1$, since $\lambda_{LR} \ll 1$).
The Toulouse point, where the Kondo problem is equivalent to a noninteracting resonant level \cite{gogolin,weiss,hewson}, occurs at $\alpha = 1$ ($g \approx 2$);
this point of highly anisotropic exchange is hardly accessible in electronic realizations of the Kondo model.
Nevertheless, the Kondo couplings still flow to the same strong-coupling fixed point as in the standard isotropic case.

The Kondo impurity is locked into a singlet with its environment at energies below the Kondo temperature $T_K$. We define it through the inverse static local impurity susceptibility,
$T_K^{-1} \equiv \partial \langle S_z \rangle/\partial B_z|_{B_z=T=0}$.
To the leading order in $I_{xy} \propto E_J^L E_J^R$ it is given by \cite{fn:anisotropy}
\begin{equation} \label{eqn:T_K}
  T_K = c(\alpha) \omega_0 \left( \frac{I_{xy}}{2 \pi a \omega_0} \right)^{2/[2 - \alpha^2]}, \qquad c(\alpha) \sim 1,
\end{equation}
with $c(0) = 1$.
For the strong-coupling physics to show up the leads should be longer than $v/T_K$ \cite{fn:lead_length}.

We now examine the ac transport properties of the circuit.
The quantum impurity causes elastic and inelastic scattering of incoming microwave photons.
The former is characterized by the elastic $T$-matrix $\hat{T}^\text{el}_{\ell^\prime | \ell}(\omega)$, defined as usual by the relation between the single photon propagators in the presence and absence of the impurity
(see SM, Sec.~SM.B \cite{sm}).
It has the structure
\begin{equation} \label{eqn:tmatrix}
  - 2 \pi i \hat{T}^\text{el}_{\ell^\prime | \ell}(\omega) = \left(
  \begin{array}{cc}
    r_L(\omega) - 1 & t_R(\omega) \\
    t_L(\omega) & r_R(\omega) - 1
  \end{array}
  \right),
\end{equation}
where $t_\ell(\omega)$ [$r_\ell(\omega)$] is the transmission [reflection] amplitude for a photon of frequency $\omega$ incoming in lead $\ell$. 

The equations of motion for the single photon propagartors allow us to derive a relation
\begin{equation} \label{eqn:t_chi}
  \hat{T}^\text{el}_{\ell^\prime | \ell}(\omega) =
  (-1)^{\delta_{\ell, \ell^\prime}-1} \omega
  \alpha_\ell \alpha_{\ell^\prime} \chi_{zz}(\omega),
\end{equation}
between all the elements of the elastic $\hat{T}$ matrix and the local dynamic differential spin susceptibility of the Kondo problem (\ref{eqn:h_k}),
$\chi_{zz}(\omega) = \dleft\langle S_z; S_z \dright\rangle_\omega$, where double angular brackets denote retarded correlators. Thus, a simple ac transport measurement on this system yields the dynamic susceptibility of the Kondo model, which is hard to access in the electronic realizations of the Kondo effect:
in those systems charge transport is weakly-coupled to the spin dynamics, whereas in our system $S_z$ is actually the electric polarization of the quantum impurity. An incoming electromagnetic wave will generate an ac voltage difference (``magnetic field'') on the ``spin''.
The impurity electric polarization will oscillate in response
[through $\chi_{zz}(\omega)$] and emit the scattered waves.

\begin{figure}[b]
\includegraphics[width=7cm,height=!]{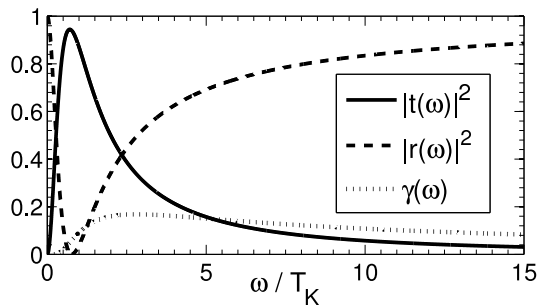}
\caption{\label{fig:scattering_probabilities}
Elastic transmission, elastic reflection, and total inelastic scattering probabilities at the Toulouse point $\alpha=1$ with left-right symmetry (hence the lead index $\ell$ was omitted) and $B_z=T=0$ (See SM, Sec.~SM.D \cite{sm}). See the text for further details.
}
\end{figure}

The frequency dependence of $\chi_{zz}$ is nonmonotonic.
We will concentrate on low temperatures ($T \ll T_K$) and small ``magnetic fields'' [cf. Eq. (4)], $B_z \ll T_K$, where Kondo physics is most clearly manifested.
The imaginary part of $\chi_{zz}(\omega)$ has a maximum at $\omega\sim T_K$
while $\text{Re} [ \chi_{zz}(\omega) ]$ alternates its sign.
These features sharpen up to 
width $\sim \alpha^2T_K$ at $\alpha\ll 1$ \cite{weiss}.
At low frequency $\omega \ll T_K$ and arbitrary $\alpha$ the susceptibility approaches a real constant,
\begin{equation} \label{eqn:chi_low_omega}
  \chi_{zz}(\omega) = \chi_0 \left( \alpha, \frac{B_z}{T_K} \right)
  \left[ 1 + i \pi \alpha^2  \omega
  \chi_0 \left( \alpha, \frac{B_z}{T_K} \right)
  \right],
\end{equation}
where $\chi_0(\alpha,B_z/T_K) \equiv \partial \langle S_z \rangle / \partial B_z$ is the static local differential susceptibility, with $\chi_0(\alpha,0) = 1/T_K$.
The coefficient of the dissipative, linear-in-frequency term, is fixed by the Shiba relation \cite{shiba75,weiss}
(See SM, Sec.~SM.C \cite{sm}).
At high frequencies, $\omega \gg T_K, B_z$, we can use perturbation theory in $I_{xy} \propto E_J^L E_J^R$ to find~%
\cite{fn:chi_toulouse}
\begin{equation} \label{eqn:chi_high_omega}
  \chi_{zz}(\omega)
  = i \frac{\pi}{4} \frac{f (\alpha)}{\omega} \left(\frac{T_K}{i \omega}\right)^{2-\alpha^2},\quad \alpha>1,
\end{equation}
where $f(\alpha) = -2 \sin(\pi \alpha^2/2) \Gamma (1 - \alpha^2) / \{\pi [c(\alpha)]^{2-\alpha^2}\}$.
At $\alpha<1$ the imaginary part of Eq.~(\ref{eqn:chi_high_omega}) still describes $\text{Im}[\chi_{zz}(\omega)]$, while the real part is dominated by another term,
$\text{Re} [ \chi_{zz}(\omega) ] \sim T_K/\omega^2$.
At $\omega\gg T_K, B_z$, the photon reflection coefficient $|r_\ell(\omega)|^2$ in the elastic channel approaches $1$, while the transmission coefficient
$| t_\ell(\omega) |^2$
scales as $(T_K/\omega)^{2(2-\alpha^2)}$ for $\alpha>1$ and as
$\alpha^4 (T_K/\omega)^{2}$ for $\alpha<1$.
The elastic scattering probabilities at the Toulouse point $\alpha=1$ are plotted in Fig.~\ref{fig:scattering_probabilities}.

Let us now turn to inelastic photon scattering.
Using Eq.~(\ref{eqn:t_chi}), the total probability of an incoming photon to be scattered inelastically is
\begin{align}
\label{eqn:gamma_tot}
  \gamma_\ell (\omega)
  & =
  1 - \left| r_\ell(\omega) \right|^2 - \left| t_\ell(\omega) \right|^2 \\ \nonumber
  & =
  4 \pi \alpha_\ell^2 \omega \text{Im} \left[ \chi_{zz}(\omega) \right] - 4 \pi^2 \alpha_\ell^2 \alpha^2 \omega^2 \left| \chi_{zz} (\omega) \right|^2.
\end{align}
This quantity would be zero for a harmonic system, but is nonzero in general
(See SM, Sec.~SM.C \cite{sm}).
Actually, for $\omega\gg T_K$ we may use Eq.~(\ref{eqn:chi_high_omega})
to find
$\gamma_\ell(\omega) \sim \alpha^4 (T_K/\omega)^{2-\alpha^2}$,
which  is parametrically larger than the elastic transmission coefficient $|t_\ell(\omega)|^2$ for any $\alpha$.
As shown in Fig.~\ref{fig:scattering_probabilities}, the total inelastic probability can reach $17\%$ at the Toulouse point $\alpha=1$, and should increase further upon increasing $\alpha$.

The measurable characteristic of the inelastic processes is the spectrum of emitted photons $\gamma_{\ell^\prime | \ell} (\omega^\prime | \omega)$, where $\gamma_{\ell^\prime | \ell} (\omega^\prime | \omega) \text{d}\omega^\prime$ is the average number of photons in the frequency interval $[\omega^\prime, \omega^\prime + \text{d}\omega^\prime]$ emitted into lead $\ell^\prime$ per each incoming photon at frequency $\omega$ in lead $\ell$ (assuming the incoming intensity is weak enough so that processes involving two or more incoming photons can be neglected). This quantity is a sum over the cross sections of all the possible multiphoton inelastic processes where one of the outgoing photons has frequency $\omega^\prime$, while integrating over all the other outgoing photons. It can also be related to local impurity correlators (See SM, Sec.~SM.D \cite{sm}).
Energy conservation leads to the relation
\begin{equation} \label{eqn:sum_rule}
  \sum_{\ell^\prime=L,R} \int_0^\infty \omega^\prime \gamma_{\ell^\prime | \ell} (\omega^\prime | \omega) \text{d}\omega^\prime = \omega \gamma_\ell(\omega).
\end{equation}

For $\omega, \omega^\prime, \omega-\omega^\prime \gg B_z, T_K$ the spectrum $\gamma_{\ell^\prime | \ell} (\omega^\prime | \omega)$ can be found perturbatively in $I_{xy} \propto E_J^L E_J^R$
(See SM, Sec.~SM.E \cite{sm}),
\begin{widetext}
\begin{multline} \label{eqn:gamma_pert0}
  \gamma_{\ell^\prime | \ell} (\omega^\prime | \omega) =
  \frac{4\pi \alpha_\ell^2 \alpha_{\ell^\prime}^2}{\omega \omega^\prime}
  \left(\frac{I_{xy}}{4 \pi a}\right)^2
  \left[
  \text{Im}
  \left[ \tilde{\chi}^\text{hl}_{+-} (\omega-\omega^\prime) \right]
  \left(
  \theta(\omega-\omega^\prime)
  \Bigl\{
  \left[ 1+n_B (\omega^\prime) \right]
  \left[ 1+n_B (\omega-\omega^\prime) \right]
  - n_B (\omega^\prime) n_B (\omega-\omega^\prime)
  \Bigr\}
  \right. \right. \\ \left. \left.
  \qquad\qquad\qquad\qquad\qquad\qquad\qquad\qquad\qquad\qquad\qquad 
  + \theta(\omega^\prime-\omega)
  \Bigl\{
  n_B (\omega^\prime) \left[ 1+n_B (\omega^\prime-\omega) \right]
  - \left[ 1+n_B (\omega^\prime)\right] n_B (\omega^\prime-\omega)
  \Bigr\}
  \right)
  \right. \\ \left.
  + \text{Im}
  \left[ \tilde{\chi}^\text{hl}_{+-} (\omega+\omega^\prime) \right]
  \Bigl\{
  n_B (\omega+\omega^\prime) \left[ 1+n_B (\omega^\prime) \right]
  - \left[ 1+n_B (\omega+\omega^\prime)\right] n_B (\omega^\prime)
  \Bigr\}
  \right],
\end{multline}
\end{widetext}
where $n_B(\omega) = 1/(e^{\omega/T}-1)$ is the Bose distribution, and $\tilde{\chi}^\text{hl}_{+-} (\omega) = \dleft\langle e^{i \alpha \tilde{\phi}_s(0)}; e^{-i \alpha \tilde{\phi}_s(0)} \dright\rangle^\text{hl}_{\omega}$,
calculated for vanishing coupling to the impurity.
The different terms in this equation account for all the possible multiphoton scattering processes. For example, the first term on the first line describes a process where an incoming photon at frequency $\omega$ is absorbed by the quantum impurity, which in turn emits a photon at frequency $\omega^\prime<\omega$ [hence the spontaneous and stimulated emission factor $1+n_B (\omega^\prime)$], plus additional photons whose energies sum up to $\omega-\omega^\prime$. It can be shown that the factors depending on $\omega-\omega^\prime$
can be written as the sum over the probabilities of distributing the energy $\omega-\omega^\prime$ among any number of photons
(See SM, Sec.~SM.E \cite{sm}).
At $T=0$ Eq.~(\ref{eqn:gamma_pert0}) yields (for $\omega^\prime<\omega$)
\begin{equation} \label{eqn:gamma_pert}
  \gamma_{\ell^\prime | \ell} (\omega^\prime | \omega) =
  \pi^2 \alpha_\ell^2 \alpha_{\ell^\prime}^2
  \tilde{f}(\alpha)
  \frac{\omega-\omega^\prime}{\omega \omega^\prime}
  \left( \frac{T_K}{\omega-\omega^\prime} \right)^{2 - \alpha^2},
\end{equation}
with $\tilde{f}(\alpha) = \sin [ \pi(\alpha^2-1)/2 ] f ( \alpha )$.
This result, together with Eqs.~(\ref{eqn:chi_high_omega})--(\ref{eqn:gamma_tot}),
obeys the sum rule (\ref{eqn:sum_rule}) to the leading order in $T_K/\omega \ll 1$.

If any of the energies $\omega$, $\omega^\prime$, or $\omega-\omega^\prime$ becomes less than $T_K$, perturbation theory in $I_{xy}$ is no longer valid. To derive the behavior of $\gamma_{\ell^\prime | \ell} (\omega^\prime | \omega)$ in these regimes, let us start from the case when all the frequencies are small, and the dynamics is governed by the strong coupling fixed point.
At low energies the impurity is screened and disappears from the problem.
According to the Nozi\`{e}res Fermi-liquid description \cite{hewson}, it leaves behind (at $B_z=0$) local scattering potential and interaction between the fermions of Eq.~(\ref{eqn:h_k}), mediated by virtual fluctuations of the Kondo impurity.
Upon bosonization, the leads are described by the first term of Eq.~(\ref{eqn:h_sb}) while the local potential and interaction acquire the form
$H_2 \sim v^2 \tilde{\rho}_s^2(0) / T_K$
\cite{fn:boundary_condition_nozieres_basis}.
This is the lowest order term allowed by symmetries;
for example, the spin density $\propto \tilde{\rho}_s(0)$ cannot appear in odd powers due to the time reversal symmetry of the Kondo model, representing the equivalence of the two impurity states in Eq.~(\ref{eqn:h_sb}) at $B_z=0$.
$H_2$ is harmonic; in order to study inelastic effects one needs to consider higher-order terms. In the absence of a magnetic field, a quartic, four-photon term
$H_4 \sim v^4 \tilde{\rho}_s^4 (0)/T_K^3$
is the lowest anharmonic term allowed, while with magnetic field three-boson scattering,
$H_3 \sim B_z v^3 \tilde{\rho}_s^3 (0)/T_K^3$,
is possible. 
Fermi's golden rule then leads to (for $\omega^\prime < \omega \ll T_K$)
\begin{equation}
\label{eqn:gamma_small_omega}
  \gamma_{\ell^\prime | \ell} (\omega^\prime | \omega) =
  \alpha_\ell^2 \alpha_{\ell^\prime}^2
  \frac{\omega \omega^\prime \left( \omega - \omega^\prime \right)
  \left[ a_B(\alpha) B_z^2 + a_\omega(\alpha) \left( \omega - \omega^\prime \right)^2 \right]}
  {T_K^6}
\end{equation}
(the coefficients $a_{B,\omega}(\alpha)$ are evaluated in the SM, Sec.~SM.F  \cite{sm} for small $\alpha$).

\begin{figure}[b]
\includegraphics[width=7cm,height=!]{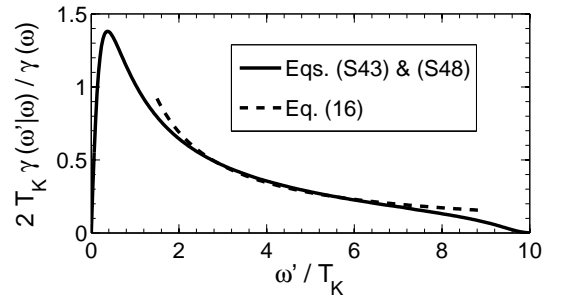}
\caption{\label{fig:spectrum_toulouse}
The inelastic spectrum normalized by the total inelastic probability at the Toulouse point $\alpha=1$ with left-right symmetry (hence the lead indices $\ell$, $\ell^\prime$ were omitted), for $\omega/T_K=10.0$, and $B_z=T=0$.
The continuous line is the exact result, see SM, Eqs.~(S43) and (S48) \cite{sm}.
The dashed line corresponds to Eq.~(\ref{eqn:gamma_pert}), valid for $\omega^\prime, \omega-\omega^\prime \gg T_K$.
See the text for further details.
The peak at $\omega^\prime \sim T_K$ sharpens, and a broad peak develops around $\omega-\omega^\prime \sim T_K$ for smaller $\alpha$;
cf.~SM, Figs.~S2 \cite{sm}.
}
\end{figure}

Returning to the high frequency regime $\omega \gg T_K$, 
the behavior near the edges of the spectrum in $\omega^\prime$ is the same as for $\omega \ll T_K$, since at $\omega \sim T_K$ a crossover, rather than a singularity, occurs.
Thus, while Eq.~(\ref{eqn:gamma_pert}) applies as long as both $\omega^\prime, \omega-\omega^\prime \gg T_K$, for small $\omega^\prime$ one has $\gamma_{\ell^\prime | \ell} (\omega^\prime | \omega) \propto \omega^\prime$, whereas for small $\omega-\omega^\prime>0$
\begin{equation} \label{eqn:gamma_small_omega_omegap}
  \gamma_{\ell^\prime | \ell} (\omega^\prime | \omega) =
  \alpha_\ell^2 \alpha_{\ell^\prime}^2
  \frac{\left( \omega-\omega^\prime \right)   \left[ a_B^{\prime}(\alpha) B_z^2 + a_\omega^\prime(\alpha) \left( \omega - \omega^\prime \right)^2 \right]}
  {\omega^2 T_K^2}
\end{equation}
(See SM, Sec.~SM.F \cite{sm}, for the small $\alpha$ values of $a^\prime_{B,\omega}(\alpha)$).
The leading dependence on $\omega-\omega^\prime$ in Eqs.~(\ref{eqn:gamma_small_omega}) and (\ref{eqn:gamma_small_omega_omegap})
changes at $B_z=0$, reflecting the higher symmetry of the system.
The resulting behavior is depicted in Fig.~\ref{fig:spectrum_toulouse} at the Toulouse point $\alpha=1$.

To conclude, we have considered the scattering of microwave photons propagating along an array of superconducting islands by a localized anharmonicity. We have shown that, contrary to the assumptions of input-output theory, linear response is typically dissipative, and inelastic scattering is therefore significant.
Photon scattering provides direct access to the dynamics of quantum impurity.
While we have concentrated on a Kondo system, these conclusions should apply to other types of quantum impurities.
Finally we note that this and related setups have been studied in the past. However, most of these works only considered equilibrium properties \cite{leclair97,camalet04,garciaripoll08}.
Elastic scattering in this system in the limit $\alpha \ll 1$ was recently studied in Ref.~\onlinecite{lehur12}. Inelastic scattering, whose probability is small in that limit (See SM, Sec.~SM.F \cite{sm}), was ignored there.

M.~G.\ would like to thank the Simons Foundation, the Fulbright Foundation, and the BIKURA (FIRST) program of the Israel Science Foundation for financial support.
M.~H.~D.\ is supported by NSF DMR grant No.~1006060 and College de France.
M.~H.\ is supported by an ANR grant (ANR-11-JS04-003-01).
L.~I.~G.\ is supported by NSF DMR Grant No.~1206612.

\renewcommand{\thesection}{SM.\Alph{section}}
\setcounter{figure}{0}
\renewcommand{\thefigure}{S\arabic{figure}}
\setcounter{equation}{0}
\renewcommand{\theequation}{S\arabic{equation}}

\newpage\pagebreak
\begin{widetext}

\section*{Supplemental Material}
In the Supplemental Material we go into some technical details of the calculations, which were omitted in the main text.

\section{Derivation of the effective low-energy Hamiltonian}

In this Section we will outline the derivation and range of validity of the effective impurity Hamiltonian,
Eqs.~(\ref{eqn:h_leads})--(\ref{eqn:bz_ejlr})
in the main text,
starting from the general array Hamiltonian,
Eq.~(\ref{eqn:h_array})
in the main text, with the parameters given in Fig.~\ref{fig:system} in the main text.
While this can be done in the general case, the resulting expressions would be quite cumbersome.
Therefore, we will concentrate on the typical regime of parameters for realistic systems \cite{Smanucharyan09}.
In particular, inter-island capacitances are typically much larger than the capacitances to the ground, and the impurity-lead capacitances are smaller than other inter-island capacitances:
$C \sim C_{LR} \gg C_L \sim C_R \gg C_g \sim C^g_L \sim C^g_R$.
As we will see in the following [cf.\ the discussion  after Eq.~(\ref{eqn:phase_shifts})], the optimal value of the impurity-lead capacitances $C_{L,R}$ is of order $\sqrt{C C_g}$, which we shall assume.
In the following we will only keep terms to the lowest nonvanishing order in the corresponding small ratios.

\subsubsection{The inverse capacitance matrix}
In order to write down the explicit form of the Hamiltonian, Eq.~(\ref{eqn:h_array}) in the main text, one needs to invert the capacitance matrix $\mathsf{C}_{m m^\prime}$, where the capacitances can be read off from Fig.~\ref{fig:system} in the main text.
This can be done similarly to the calculation of the Green functions of a noninteracting tight binding model \cite{Seconomou}, where the capacitances to the ground take the place of the onsite energies, and the inter-island capacitances are analogous to the hopping matrix elements.
The presence of large inter-island capacitances makes the inverse capacitance matrix long-ranged: for a uniform lead one has [$\kappa = 1 + C_g / (2 C)$]
\begin{equation} \label{eqn:cinv_0}
  \left[ \mathsf{C}^{-1} \right]^{(0)}_{m m^\prime}
  =
  \frac{1}{2 C \sqrt{\kappa^2-1}}
  \left( \kappa - \sqrt{\kappa^2 - 1} \right)^{|m - m^\prime|}
  \sim
  \frac{1}{2 \sqrt{C C_g}} \left( 1 - \sqrt{\frac{C_g}{C}} \right)^{|m - m^\prime|},
\end{equation}
whereas for a half-infinite lead ($\text{hl}$),
\begin{align} \label{eqn:cinv_hl}
  \left[ \mathsf{C}^{-1} \right]^\text{hl}_{m m^\prime}
  = &
  \frac{1}{2 C \sqrt{\kappa^2-1}}
  \left[
  \left( \kappa - \sqrt{\kappa^2 - 1} \right)^{|m - m^\prime|} +
  \left( \kappa - \sqrt{\kappa^2 - 1} \right)^{|m + m^\prime - 1|}
  \right] \nonumber \\
  \sim &
  \frac{1}{2 \sqrt{C C_g}}
  \left[
  \left( 1 - \sqrt{\frac{C_g}{C}} \right)^{|m - m^\prime|} +
  \left( 1 - \sqrt{\frac{C_g}{C}} \right)^{|m + m^\prime - 1|}
  \right],
\end{align}
with $m,m^\prime>0$.

We can now write down the elements of the inverse capacitance matrix of the system in the presence of the quantum impurity, which appear in the first term of the Hamiltonian, Eq.~(\ref{eqn:h_array}) in the main text.
The impurity sub-block of the inverse capacitance matrix is given by
\begin{align} \label{eqn:cinv_imp}
  \left(
  \begin{matrix}
  \left[ \mathsf{C}^{-1} \right]_{LL} & \left[ \mathsf{C}^{-1} \right]_{LR} \\
  \left[ \mathsf{C}^{-1} \right]_{RL} & \left[ \mathsf{C}^{-1} \right]_{RR}
  \end{matrix}
  \right)
  = &
  \left(
  \begin{matrix}
  C^g_L + C_L + C_{LR}
  - \left(C_{L}\right)^2 \left[ \mathsf{C}^{-1} \right]^{\text{hl},L}_{1,1} &
  - C_{LR} \\
  - C_{LR} &
  C^g_R + C_R + C_{LR}
  - \left(C_{R}\right)^2 \left[ \mathsf{C}^{-1} \right]^{\text{hl},R}_{1,1}
  \end{matrix}
  \right)^{-1}
  \nonumber \\
  \sim &
  \frac{1}{\tilde{C}_L+\tilde{C}_R}
  \left(
  \begin{matrix}
    1 & 1 \\ 1 & 1
  \end{matrix}
  \right)
  +\frac{1}{C_{LR} (\tilde{C}_L+\tilde{C}_R)^2}
  \left(
  \begin{matrix}
     \tilde{C}^{2}_R & - \tilde{C}_L \tilde{C}_R \\ -\tilde{C}_L \tilde{C}_R & \tilde{C}^{2}_L
  \end{matrix}
  \right),
\end{align}
where $\tilde{C}_\ell = C_\ell \sqrt{C C_g} / (C_\ell + \sqrt{C C_g})$, and
\begin{equation} \label{eqn:cinv_lead_aux}
  \left[ \mathsf{C}^{-1} \right]^{\text{hl},\ell}_{m m^\prime} =
  \left[ \mathsf{C}^{-1} \right]^{\text{hl}}_{m m^\prime} -
  \left[ \mathsf{C}^{-1} \right]^{\text{hl}}_{m 1}
  \frac{C_\ell}{1 + C_\ell \left[ \mathsf{C}^{-1} \right]^{\text{hl}}_{1 1}}
  \left[ \mathsf{C}^{-1} \right]^{\text{hl}}_{1 m^\prime},
\end{equation}
are the elements of the inverse capacitance matrix of a half-infinite lead terminated by a capacitance to the ground whose magnitude is $C_\ell$.
The first term in the last line of Eq.~(\ref{eqn:cinv_imp}) dominates the dynamics of the impurity total charge, while the second governs the behavior of its polarization.

The impurity-leads elements of the inverse capacitance matrix appearing in Eq.~(\ref{eqn:h_array}) in the main text are given by
\begin{equation} \label{eqn:cinv_imp_lead}
  \left[ \mathsf{C}^{-1} \right]_{\ell, m^\prime \ell^\prime} =
  \left[ \mathsf{C}^{-1} \right]_{\ell,\ell^\prime} C_{\ell^\prime}
  \left[\mathsf{C}^{-1}\right]^{\text{hl},\ell^\prime}_{1, m^\prime},
\end{equation}
whereas the lead-lead elements are modified to (in the following subsection we treat the lead dynamics in a Lagrangian formulation, and thus do not use this formula; it is given here for reference):
\begin{equation} \label{eqn:cinv_lead}
  \left[ \mathsf{C}^{-1} \right]_{m \ell, m^\prime \ell^\prime} =
  \delta_{\ell, \ell^\prime}
  \left[ \mathsf{C}^{-1} \right]^{\text{hl},\ell}_{m, m^\prime}
  + \left[ \mathsf{C}^{-1} \right]^{\text{hl},\ell}_{m, 1}
  C_\ell
  \left[ \mathsf{C}^{-1} \right]^{-1}_{\ell, \ell^\prime}
  C_{\ell^\prime}
  \left[ \mathsf{C}^{-1} \right]^{\text{hl},\ell^\prime}_{1, m^\prime}.
\end{equation}

\subsubsection{Validity of the low-energy Hamiltonian of the leads~(\ref{eqn:h_leads})}
As Eqs.~(\ref{eqn:cinv_0})--(\ref{eqn:cinv_hl}) show, the large inter-island capacitances result in a long range of the inverse capacitance matrix.
For a uniform array this may be ignored as long as one is interested in modes with wavelengths longer than the charge screening length $a \sqrt{C/C_g}$ [see also Eq.~(\ref{eqn:dispersion}) below], leading to the low-energy effective leads Hamiltonian, Eq.~(\ref{eqn:h_leads}) in the main text.
However, the situation is more complicated in the presence of the nonuniformity created by the quantum impurity.
The capacitive coupling to the impurity modifies the dynamics of the leads electromagnetic modes, allowing for their scattering and transmission between left and right even for $E_J^L = E_J^R = 0$, i.e., in the absence of the quantum impurity dynamics. In this subsection we will show that these effects can still be ignored, and Eq.~(\ref{eqn:h_leads}) in the main text may still be used, at energies lower than $\omega_0 \sim (v/a) \sqrt{C_g/C}$.

Let us therefore examine the case $E_J^L = E_J^R = 0$, assuming further right-left symmetry $C^g_L=C^g_R=C^g_0$, $C_L=C_R=C_0$ (effects of right-left asymmetry will be discussed below). After replacing the Josephson couplings by quadratic terms, as appropriate for $E_J C/(2e)^2 \gg 1$, the symmetric and antisymmetric modes with respect to the center of the array decouple. Relabeling the islands to the right/left of the impurity by $m = \pm 1, \pm 2, \cdots$, respectively, the symmetric and antisymmetric modes are defined by
\begin{align}
   \phi^{\pm}_m &= \frac{\phi_m \pm \phi_{-m}}{\sqrt{2}}, \qquad m>0 \\
   \phi^{\pm}_0 &= \frac{\phi_R \pm \phi_L}{\sqrt{2}},
\end{align}
and similarly for the operators $n^{\pm}_m$. Their dynamics is governed by the Lagrangian
\begin{equation}
  \mathcal{L}_0^\pm = \frac{1}{2} \sum_{m>0} \left[
  C_g \left( \frac{\dot{\phi}^\pm_m}{2e} \right)^2
  + C \left( \frac{\dot{\phi}^\pm_m - \dot{\phi}^\pm_{m+1}}{2e} \right)^2
  - E_J \left( \phi^\pm_m - \phi^\pm_{m+1} \right)^2
  \right]
  + \frac{1}{2} C^{g \pm}_0 \left( \frac{\dot{\phi}^\pm_0}{2e} \right)^2
  + \frac{1}{2} C_0 \left( \frac{\dot{\phi}^\pm_0 - \dot{\phi}^\pm_{1}}{2e} \right)^2
\end{equation}
where $C_0^{g +} = C^g_0$, $C_0^{g -} = 2 C_{LR} + C^g_0$.
The eigenfrequencies of the system are then
\begin{equation} \label{eqn:dispersion}
  \omega(k) = 2 \sqrt{\frac{(2e)^2 E_J}{C_g}}
  \frac{\sin (k a/2)}{\sqrt{1 + 4\frac{C}{C_g}\sin^2(k a/2)}},
\end{equation}
while the eigenmode expansion is:
\begin{equation}
  \phi^\pm_m = \sqrt{\frac{2}{L}} \sum_k \phi^\pm_k \cos \left[ k a \left( m - \frac{1}{2} \right) - \delta^\pm_k \right],
  \qquad m>0,
\end{equation}
where $L$ is the leads length (the allowed values of $k$ depend on the exact boundary conditions at the far end of the lead, but this is immaterial for the quantum impurity dynamics we are after), and the scattering phase of the eigenmodes is given by
\begin{equation} \label{eqn:phase_shifts}
  e^{2 i (\delta^\pm_k - k a)} =
  \frac{2 \left[ (2e)^2 E_J - \omega^2(k) C \right] \sin(k a/2) - i \omega^2(k) C_1^{g \pm} e^{-i k a/2}}
  {2 \left[ (2e)^2 E_J - \omega^2(k) C \right] \sin(k a/2) + i \omega^2(k) C_1^{g \pm} e^{i k a/2}}
\end{equation}
where $C_{1}^{g \pm} = C_g + C_0^{g \pm} C_0 / (C_0^{g \pm} + C_0)$ is the effective total ground capacitance of the island $m=1$. Therefore, $|\delta^-_k| > |\delta^+_k|$. The phase shifts are negligible for $k a \ll (2e)^2 E_J / (v^2 C_1^{g -}) \sim C_g/C_0$.

As a result of the above, the use of the low energy effective Hamiltonian, Eqs.~(\ref{eqn:h_leads})--(\ref{eqn:h_lm}) in the main text, as well as neglecting of scattering of photons by the impurity capacitances, are justified only at frequencies smaller than $\max[ (v/a) \sqrt{C_g/C}, (v/a) C_g/C_0]$. Choosing $C_0$ of the order of $\sqrt{C C_g}$ is optimal, as mentioned above, in the sense of matching the two cutoffs and thus not ``wasting'' frequency range.
In this low frequency limit the eigenmode expansion of the occupancies $n_m^\pm = \partial \mathcal{L}_0^\pm/\partial \dot{\phi}_m^\pm$ takes the form
\begin{align}
  \label{eqn:nm_mode}
  n^\pm_m & = \sqrt{\frac{2}{L}} \sum_k n_k^\pm \cos \left[ k a \left( m - \frac{1}{2} \right) \right],
  \qquad m>1, \\
  \label{eqn:n1_mode}
  n^\pm_1 & = \frac{C^{g \pm}_1}{C_g} \sqrt{\frac{2}{L}} \sum_k n_k^\pm \cos(k a/2),
\end{align}
i.e., only the behavior at $m=1$ is significantly affected by the inter-island capacitances.
If we lift the restriction of right-left symmetry, a similar calculation shows that at frequencies much smaller than $\max[ (v/a) \sqrt{C_g/C}, (v/a) C_g/C_{L,R}]$ all the above essentially remains the same, except that $C_0$ is replaced by $2 C_L C_R / (C_L + C_R)$ [cf.\ Eq.~(\ref{eqn:cinv_imp_rho}) below].

\subsubsection{The quantum impurity Hamiltonian}
Building on the basis laid down in the previous subsections, we will now write down the quantum impurity part of the Hamiltonian, Eq.~(\ref{eqn:h_array}) in the main text, at frequencies smaller than $\omega_0 \sim (v/a) \sqrt{C_g/C}$.
Let us start from the charging part. The inter-impurity capacitive coupling is given by Eq.~(\ref{eqn:cinv_imp}).
Using Eqs.~(\ref{eqn:cinv_hl})--(\ref{eqn:cinv_imp_lead}) and (\ref{eqn:nm_mode})--(\ref{eqn:n1_mode}), the effective impurity-lead capacitive coupling at low frequencies assumes the form
\begin{multline} \label{eqn:cinv_imp_rho}
  \frac{(2 e)^2}{\sqrt{2}}
  \sum_{\ell,m>0} \left( n_\ell - \frac{C^g_\ell V^g_\ell}{2 e} \right)
  \left\{
  \left( \left[ \mathsf{C}^{-1} \right]_{\ell, m L} + \left[ \mathsf{C}^{-1} \right]_{\ell, m R} \right) n_m^+
  + \left( \left[ \mathsf{C}^{-1} \right]_{\ell, m L} - \left[ \mathsf{C}^{-1} \right]_{\ell, m R} \right) n_m^-
  \right\}
  \\
  \sim
  \frac{(2 e)^2 \tilde{C}_L \tilde{C}_R}{C_g C_{LR} (\tilde{C}_L + \tilde{C}_R)^2}
  \left( 1 + \frac{2 C_L C_R}{(C_L + C_R) \sqrt{C C_g}} \right)
  \left[ \tilde{C}_L \left( n_L - \frac{C^g_L V^g_L}{2 e} \right) - \tilde{C}_R \left( n_R - \frac{C^g_R V^g_R}{2 e} \right) \right]
  \left[ \rho_L(0) - \rho_R(0) \right]
\end{multline}
where
\begin{equation}
  \label{eqn:rho_mode}
  \rho_{L,R}(x) = \frac{1}{\sqrt{L}} \sum_k (n_k^+ \mp n_k^-) \cos(k x),
\end{equation}
are the fields occurring in the continuum version of the lead Hamiltonian, Eq.~(\ref{eqn:h_leads}) in the main text.

Combining Eqs.~(\ref{eqn:cinv_imp}) and (\ref{eqn:cinv_imp_rho}) together with the impurity-lead Josephson coupling,
the impurity Hamiltonian assumes the form
\begin{equation}
\begin{split}
\label{eqn:h_imp}
H_\text{imp} = &
\frac{(2e)^2}{2(\tilde{C}_L+\tilde{C}_R)}
\left[ n_L + n_R
- \frac{C^g_{L} V^g_{L} + C^g_{R} V^g_{R}}{2e} 
\right]^2
+\frac{(2e)^2}{2 C_{LR} (\tilde{C}_L + \tilde{C}_R)^2}
\left[
\tilde{C}_R n_L - \tilde{C}_L n_R
- \frac{\tilde{C}_R C^g_L V^g_L - \tilde{C}_L C^g_R V^g_R}{2e}
\right]^2
\\ &
  + \frac{(2 e)^2 \tilde{C}_L \tilde{C}_R}{C_g C_{LR} (\tilde{C}_L + \tilde{C}_R)^2}
  \left( 1 + \frac{2 C_L C_R}{(C_L + C_R) \sqrt{C C_g}} \right)
  \left[ \tilde{C}_L \left( n_L - \frac{C^g_L V^g_L}{2 e} \right) - \tilde{C}_R \left( n_R - \frac{C^g_R V^g_R}{2 e} \right) \right]
  \left[ \rho_L(0) - \rho_R(0) \right]
\\ &
-E_J^L \cos \left[ \varphi_L-\phi_L(0)\right]
-E_J^R \cos \left[ \varphi_R-\phi_R(0)\right].
\end{split}
\end{equation}
Here $n_L$, $\varphi_L$ and $n_R$, $\varphi_R$ are the number and phase operators of the islands $L$ and $R$,  respectively.

\subsubsection{Derivation and validity of the effective spin impurity Hamiltonian~(\ref{eqn:h_lm})}

We will now outline how the effective spin impurity Hamiltonian, Eqs.~(\ref{eqn:h_lm})--(\ref{eqn:bz_ejlr}) in the main text, can be derived from the more general form~(\ref{eqn:h_imp}) under suitable conditions.

As mentioned in the main text, the quantum dynamics of phases $\varphi_{L,R}$ strongly depends on the ratio of the Josephson energies $E_J^{L,R}$ to the charging energy $E_C^\text{imp}=(2e)^2/[2 (\tilde{C}_L+\tilde{C}_R)]$.
If the latter is small, phase fluctuations are small and one may expand the Josephson energy part of $H_\text{imp}$ to second order in the respective arguments. The resulting harmonic version of $H_\text{imp}$ would lead to elastic photon scattering only.
In the opposite limit, $E_C^\text{imp}\gg E_J^{L,R}$, the total charge $- 2e (n_L+n_R)$ of the two islands is fixed by the large Coulomb energy penalty. If the gate voltages are tuned to a total charge of a single Cooper pair, $(C^g_L V^g_L + C^g_R V^g_R)/(2e)=1$, then $n_L+n_R=1$. When $E_J^{L,R}$ are zero, the charge of each of the islands can only take the integer values $0$ or $1$, and does not vary in time.
The possible occupancy states are thus
$| 0_L,1_R \rangle$ (i.e., $n_L=0$, $n_R=1$) and
$| 1_L,0_R \rangle$ (i.e., $n_L=1$, $n_R=0$).
We label these two charge configurations by the states of a pseudospin, $S_z = (n_L-n_R)/2 = \pm 1/2$, so that
$S_+ = | 1_L,0_R \rangle \langle 0_L,1_R |$, $S_-=(S_+)^\dagger$. Hence, $S_z$ and $S_\pm$ obey the standard spin commutation relations.

Finite $E_J^L$ and $E_J^R$ allow for switching between the configurations
$| 0_L,1_R \rangle$ and $| 1_L,0_R \rangle$ (i.e., flipping of the pseudospin) by virtual transitions to states with $n_L+n_R \ne 1$, with energies higher by $\sim E_C^\text{imp}$ (which is of the order of $\omega_0 \sim (v/a) \sqrt{C_g/C}$ for $E_J \gtrsim E_{C_g}$).
The charge part of the Hamiltonian~(\ref{eqn:h_imp}) can be projected into the low energy sector by substituting $n_{L,R} = 1/2 \pm S_z$, yielding the last two terms of Eq.~(\ref{eqn:h_lm}) in the main text. Terms that do not involve the impurity degrees of freedom can be gauged out up to a renormalization of the magnetic field $B_z$ [corresponding to the term proportional to $\lambda_{LR}^2$ in Eq.~(\ref{eqn:bz_ejlr}) of the main text].

As for the Josephson part of Eq.~(\ref{eqn:h_imp}), one may
perform a Schrieffer-Wolff transformation \cite{Shewson} in order to account for processes involving high-energy virtual states. This results in the first term of Eq.~(\ref{eqn:h_lm}) in the main text.
Here it should be noted that the Schrieffer-Wolff transformation also yields terms containing $S_z \rho_{L,R}(0)$. These would have amplitudes $\sim (E_J^\ell)^2/(g E_C^\text{imp})$. They are thus small compared to the ones of the same structure in Eqs.~(\ref{eqn:h_lm}) and (\ref{eqn:h_k}) in the main text by the factor $\sim (E_J^\ell)^2/[E_{C_g} E_C^\text{imp} (1 - \alpha/\sqrt{2})]$ and can be neglected, unless one is in the vicinity of the isotropic Kondo model, $\alpha^2 \approx 2$.

\section{Different formulations of elastic scattering}

In this section we will examine different formulations of the elastic scattering problem in the system, and demonstrate their equivalence.
One approach, alluded to in the discussion of Eq.~(\ref{eqn:tmatrix}) in the main text, is to look at the single photon elastic scattering coefficients.
These are encapsulated in the behavior of the time-ordered single-photon Green function
$\mathcal{G}_{\ell^\prime | \ell} (x^\prime | x; \omega)$
(with $\ell, \ell^\prime = L, R$), where
$\mathcal{G}_{\ell^\prime | \ell} (x^\prime | x; t) \equiv -i \langle \hat{\mathcal{T}} \rho (x^\prime,t) \rho(x,0) \rangle$,
$\hat{\mathcal{T}}$ being the time-ordering operator
\cite{Sfn:scatter_phi}.
This Green function is related to the corresponding propagator
$\delta_{\ell \ell^\prime} \mathcal{G}^\text{hl} (x^\prime | x; \omega)$ for a half-infinite lead detached from the impurity ($E_J^{L,R} = 0$, $C_{L,R} = 0$) by
\begin{equation}
\label{eqn:gphoton}
  \mathcal{G}_{\ell^\prime | \ell} (x^\prime | x; \omega) =
  \delta_{\ell \ell^\prime} \mathcal{G}^\text{hl} (x^\prime | x; \omega)
  -
  \mathcal{G}^\text{hl} (x^\prime | 0; \omega)
  \frac{\pi v^2}{g \omega} \hat{T}^\text{el}_{\ell^\prime | \ell}(\omega)
  \mathcal{G}^\text{hl} (0 | x; \omega),
\end{equation}
where the elastic $T$-matrix $\hat{T}^\text{el}_{\ell^\prime | \ell} (\omega)$ has the structure given by Eq.~(\ref{eqn:tmatrix}) in the main text. The elastic $T$-matrix appears with a prefactor $\pi v^2/(g \omega)$ in Eq.~(\ref{eqn:gphoton}) to compensate for the prefactors in the expansion of $\rho_\ell(x)$ in terms of the photon creation and annihilation operators [eigenmodes of the lead Hamiltonian, Eq.~(\ref{eqn:h_leads}) in the main text], which reads
(for a lead of length $L$ with no-current boundary condition, $\partial_x \tilde{\phi}_s(0) = 0$, when decoupled from the impurity)
\begin{equation}
  \label{eqn:rho}
  \rho_s(x) = \sum_{\substack{q = \pi n/L,\\ q>0}} i \sqrt{\frac{q g}{\pi L}} \cos(q x)
  \left( a_{s,q} - a^\dagger_{s,q} \right).
\end{equation}

There is another way to look at elastic scattering, which is equivalent to the previous one
at zero temperature and can serve as its generalization to nonzero temperatures: one may add to the Hamiltonian, Eqs.~(2)--(3) in the main text, 
the term
$H_{ac} = 2 V_0 \cos(\omega t) \rho_L (x_\text{in})$,
describing an ac gate voltage coupled to the island at $x_\text{in}$ in the left lead. This perturbation generates waves propagating to the left and to the right. The latter will scatter off the quantum impurity. The transmission amplitude $t_L(\omega)$ \cite{Sfn:amplitudes}
for waves coming from the left is then the ratio of the transmitted and incoming average currents,
\begin{equation} \label{eqn:t_l}
  t_L (\omega) = \frac{ I_\text{trans} (\omega)}{I_\text{in} (\omega)} = \frac
  {\dleft\langle \partial_x \phi_R(x_\text{out}); \rho_L(x_\text{in}) \dright\rangle_\omega}
  {\dleft\langle \partial_x \phi_R(x_\text{out}); \rho_L(x_\text{in}) \dright\rangle_\omega^{(0)}},
\end{equation}
where transmitted current is measured at $x_\text{out}$ in the right lead.
In the second equality we have written the transmission coefficient as the ratio of the conductance of the system with impurity (double angular brackets denote retarded correlation functions) and the corresponding quantity for a uniform array.
Thus, $|t_L(\omega)| \le 1$ and is independent of $x_\text{in,out}$, as required.
We can write down similar expressions for the other scattering amplitudes.

To show the equivalence of these two formulations, we start from Eqs.~(\ref{eqn:gphoton})--(\ref{eqn:rho}), and note that
at zero temperature time-ordered and retarded Green functions are the same for positive frequencies, $\omega>0$. Therefore, Eq.~(\ref{eqn:gphoton}) yields for the transmission coefficient $t_L(\omega)$ for photons coming from the left,
\begin{equation} \label{eqn:t_l_photon}
  t_L (\omega)
  =
  2 \pi i
  \frac{g \omega}{\pi v^2}
  \frac{\dleft\langle \rho_R(x^\prime), \rho_L(x) \dright\rangle_\omega}
  {\dleft\langle \rho_R(x^\prime), \rho_R(0) \dright\rangle^\text{hl}_\omega
  \dleft\langle \rho_L(0), \rho_L(x) \dright\rangle^\text{hl}_\omega}.
\end{equation}
Now, for the lead Hamiltonian, Eq.~(\ref{eqn:h_leads}) in the main text, one has \cite{Sgogolin}
\begin{align}
  \dleft\langle \rho_\ell(x), \rho_\ell(0) \dright\rangle^\text{hl}_\omega
  = &
  \frac{i g \omega}{\pi v^2} e^{i \omega x / v},
  \nonumber \\
  \dleft\langle \rho_R(x_R), \rho_L(x_L) \dright\rangle^{(0)}_\omega
  = &
  \frac{i g \omega}{2 \pi v^2} e^{i \omega (x_R-x_L) / v},
\end{align}
where the superscript $(0)$ denotes the propagator for a uniform waveguide, with no quantum impurities, as in the main text.
Thus, Eq.~(\ref{eqn:t_l_photon}) can be rewritten as:
\begin{equation}
  t_L (\omega)
  =
  \frac{\dleft\langle \rho_R(x^\prime), \rho_L(x) \dright\rangle_\omega}
  {\dleft\langle \rho_R(x^\prime), \rho_L(x) \dright\rangle^{(0)}_\omega}
  =
  \frac{\dleft\langle \partial_x \phi_R(x^\prime), \rho_L(x) \dright\rangle_\omega}
  {\dleft\langle \partial_x \phi_R(x^\prime), \rho_L(x) \dright\rangle^{(0)}_\omega},
\end{equation}
where the last equality results from the equation of motion
$\partial_t \rho_\ell(x,t) = (v g/\pi) \partial_x^2 \phi_\ell(x,t)$.
We have thus proven the equivalence of Eq.~(\ref{eqn:gphoton}) 
with Eq.~(\ref{eqn:t_l}) at zero temperature.
The latter equation can thus be thought of as a generalization of the former one to nonzero temperatures.
Similar treatment applies to the other elastic scattering coefficients.

\section{Shiba Relations from Photon Scattering}
The Shiba relation connects the low frequency behavior of the real and imaginary part of the Kondo local spin susceptibility $\chi_{zz}(\omega)$ \cite{Sshiba75,Sweiss,Shewson}.
In this Section we will show how considerations based on photon-scattering can be used to rederive, as well as to generalize, this relation.
Expanding the local spin susceptibility in powers of $\omega$,
\begin{align}
 \text{Re}[\chi_{zz}(\omega)] = & \chi_0 + \chi_2 \omega^2 + \cdots, \\
 \text{Im}[\chi_{zz}(\omega)] = & \chi_1 \omega + \chi_3 \omega^3 + \cdots,
\end{align}
and substituting in Eq.~(\ref{eqn:gamma_tot}) in the main text, we obtain an expansion of the total inelastic scattering probability $\gamma_\ell(\omega)$ in powers of $\omega \ll T_K$, with coefficients depending on the $\chi_i$.
On the other hand, from Eq.~(\ref{eqn:gamma_small_omega}) and Eq.~(\ref{eqn:sum_rule}) in the main text it follows that when $\omega$ is small, $\gamma_\ell(\omega) \sim \omega^4$ in the presence of a magnetic field, while $\gamma_\ell(\omega) \sim \omega^6$ for $B_z = 0$.
Comparing these results we find
that the vanishing of the total inelastic scattering probability  $\gamma_\ell(\omega)$ to order $\omega^2$ leads to the Shiba relation [cf.\ Eq.~(\ref{eqn:chi_low_omega}) in the main text] \cite{Sweiss}:
\begin{equation}
  \chi_1 = \pi \alpha^2 (\chi_0)^2,
\end{equation}
whereas the vanishing of $\gamma_\ell(\omega)$ to order $\omega^4$ in the absence of a magnetic field leads to a new, higher order, Shiba-like relation:
\begin{equation}
  \chi_3 = \pi \alpha^2 [2 \chi_0 \chi_2 + (\chi_1)^2].
\end{equation}
This latter relation can be easily verified to hold at the exactly-solvable Toulouse point $\alpha=1$ \cite{Sgogolin,Sweiss,Shewson}, where the susceptibility is given by Eq.~(\ref{eqn:chi_toulouse}) below.

\section{Inelastic spectrum from nonlinear response functions}
The inelastic spectrum $\gamma_{\ell^\prime \vert \ell} (\omega^\prime \vert \omega) \text{d}\omega^\prime$ is defined in the main text as the average number
of photons within a frequency interval $\text{d}\omega^\prime$ around
$\omega^\prime$ emitted through lead $\ell^\prime$ for each incoming photons at frequency $\omega$ in lead $\ell$. Thus, it is a sum over the partial cross sections for all the possible multiphoton scattering processes, integrated over all the photons except the one with frequency $\omega^\prime$.
In this Section we will show that, similarly to 
Eq.~(\ref{eqn:t_chi}) in the main text for elastic scattering,
$\gamma_{\ell^\prime \vert \ell} (\omega^\prime \vert \omega)$ can also be expressed in terms of response functions, and related to local spin correlators.
Since the number of photons emitted at frequency $\omega^\prime$ is proportional to the flux of incoming photons, or incoming energy flux (assuming scattering between two or more incoming photons is negligible), we need to consider \emph{second} order response to the ac source of incoming photons.

To spare us the need to carry around the indices $\ell$, $\ell^\prime$ in the following, we define
\begin{equation} \label{eqn:gamma_ells}
  \gamma_s (\omega^\prime \vert \omega) \equiv \sum_{\ell,\ell^\prime} \gamma_{\ell^\prime \vert \ell} (\omega^\prime \vert \omega),
  \quad
  \gamma_{\ell^\prime \vert \ell} (\omega^\prime \vert \omega) =
  \frac{\alpha_\ell^2 \alpha_{\ell^\prime}^2}{\alpha^4}
  \gamma_s (\omega^\prime \vert \omega),
\end{equation}
The second relation results from the fact that only the ``spin fields'' $\tilde{\phi}_s(x)$ and $\tilde{\rho}_s(x)$ are coupled to the impurity [cf.\ the discussion following Eq.~(\ref{eqn:h_lm}) in the main text].

The quantity of interest here is the time-averaged rate of change of the photon number
$n_{s,k^\prime} = \tilde{a}^\dagger_{s,k^\prime} \tilde{a}_{s,k^\prime}$, $k^\prime = \omega^\prime/v$,
to second order in an applied ac voltage $H_{ac} =  V(t) e^{\eta t} \tilde{\rho}_s (x_{in})$, with $V(t) = 2 V_0 \cos (\omega t)$, $\eta \to 0^+$, divided by the incoming flux of photons of frequency $\omega$
\cite{Sfn:V_rhos}.
The photon creation and annihilation operators $\tilde{a}^\dagger_{s,q}$ and $\tilde{a}_{s,q}$ are the Fourier modes of the bosonic fields (obeying the no-current boundary condition $\partial_x \tilde{\phi}_s(0) = 0$ when decoupled from the impurity),
\begin{align}
  \label{eqn:phi_s}
  \tilde{\phi}_s(x) = & \sum_{\substack{q = \pi n/L,\\ q>0}} \sqrt{\frac{\pi}{q L}} \cos(q x) \left( \tilde{a}_{s,q} + \tilde{a}^\dagger_{s,q} \right)
  \\
  \label{eqn:rho_s}
  \tilde{\rho}_s(x) = & \sum_{\substack{q = \pi n/L,\\ q>0}} i \sqrt{\frac{q}{\pi L}} \cos(q x)
  \left( \tilde{a}_{s,q} - \tilde{a}^\dagger_{s,q} \right)
\end{align}
where $L$ is the lead length.
The second order Kubo formula reads
\begin{equation}
\left\langle \tilde{n}_{s,k^\prime}(t) \right\rangle^{(2)} =
\frac{1}{2}
\int_{-\infty}^\infty \text{d}t^\prime \int_{-\infty}^\infty \text{d}t^{\prime \prime}
G^{cqq}_{\tilde{n}_{s,k^\prime}; \tilde{\rho}_s (x_{in}); \tilde{\rho}_s (x_{in})} (t-t^\prime,t-t^{\prime \prime})
V(t^\prime) V(t^{\prime \prime}),
\end{equation}
where the second order response function is:
\begin{equation}
G^{cqq}_{\tilde{n}_{s,k^\prime}; \tilde{\rho}_s (x_{in}); \tilde{\rho}_s (x_{in})} (t-t^\prime,t-t^{\prime \prime}) =
- \theta(t-t^\prime)\theta(t^\prime-t^{\prime \prime})
\left\langle
\left[ \left[ \tilde{n}_{s,k^\prime}(t),
\tilde{\rho}_s (x_{in}, t^\prime) \right],
\tilde{\rho}_s (x_{in}, t^{\prime \prime}) \right]
\right\rangle
+\left\{ t^\prime \leftrightarrow t^{\prime \prime} \right\}.
\end{equation}
Here $c$, $q$ denote ``classical'' and ``quantum'' fields in the Keldysh formalism \cite{Skamenev,Schou85}, i.e., the sum and difference, respectively, of fields on the forward and backward contours.
The time-average of the photon production \emph{rate} can thus be written as:
\begin{equation}
\overline{\frac{\text{d} \left\langle \tilde{n}_{s,k^\prime} \right\rangle^{(2)}}{\text{d}t}} =
2 \eta G^{cqq}_{\tilde{n}_{s,k^\prime}; \tilde{\rho}_s (x_{in}); \tilde{\rho}_s (x_{in})} (\omega+i\eta,-\omega+i\eta) |V_0|^2,
\end{equation}
where frequency arguments are in correspondence with time arguments in the previous equation.
Since we should take the limit $\eta \to 0^+$, the factor of
$\eta$ in this formula implies that we should be looking
for contributions to the correlation function which are singular in that limit.
Multiplying $\overline{\text{d} \langle \tilde{n}_{s,k^\prime} \rangle^{(2)} / \text{d}t}$ by the photon density of states $L/(\pi v)$, and dividing by $\omega |V_0|^2 / (\pi v^2)$, the rate of creation of photons with frequency $\omega$ moving towards the impurity by the source $V(t)$, we have
\begin{equation} \label{eqn:2nd_3pt}
  \gamma_s(\omega^\prime | \omega) =
  \frac{2 \eta v L}{\omega}
  G^{cqq}_{\tilde{n}_{s,k^\prime}; \tilde{\rho}_s (x_{in}); \tilde{\rho}_s (x_{in})} (\omega+i\eta,-\omega+i\eta),
\end{equation}

For subsequent calculations it is better to look at a more general correlation function, where $\tilde{a}^\dagger_{s,k^\prime}$ and $\tilde{a}_{s,k^\prime}$ have different time arguments $t_1$ and $t_2$, respectively, and take the limit of $t_1 = t_2 = t$ only at the
end. A suitable correlator is the
following Green function, which appears naturally in the
Keldysh formalism \cite{Schou85}:
\begin{multline} \label{eqn:gccqq}
G^{ccqq}_{\tilde{a}^\dagger_{s,k^\prime}; \tilde{a}_{s,k^\prime}; \tilde{\rho}_s (x_{in}); \tilde{\rho}_s (x_{in})} (t_1-t_2, t_1-t^\prime,t_2-t^{\prime \prime}) = \\
\quad
i \theta(t_1-t_2) \theta(t_2-t^\prime)\theta(t^\prime-t^{\prime \prime})
\left\langle
\left[ \left[ \left\{ \tilde{a}^\dagger_{s,k^\prime}(t_1), \tilde{a}_{s,k^\prime}(t_2)\right\},
\tilde{\rho}_s (x_{in}, t^\prime) \right],
\tilde{\rho}_s (x_{in}, t^{\prime\prime}) \right]
\right\rangle \\
+ i \theta(t_1-t^\prime) \theta(t^\prime-t_2) \theta(t_2-t^{\prime \prime})
\left\langle
\left[ \left\{ \left[ \tilde{a}^\dagger_{s,k^\prime}(t_1),
\tilde{\rho}_s (x_{in}, t^\prime) \right],
\tilde{a}_{s,k^\prime}(t_2) \right\},
\tilde{\rho}_s (x_{in}, t^{\prime\prime}) \right]
\right\rangle \\
+ i \theta(t_1-t^\prime) \theta(t^\prime-t^{\prime \prime}) \theta(t^{\prime \prime} - t_2)
\left\langle
\left\{ \left[ \left[ \tilde{a}^\dagger_{s,k^\prime}(t_1),
\tilde{\rho}_s (x_{in}, t^\prime) \right],
\tilde{\rho}_s (x_{in}, t^{\prime\prime}) \right],
\tilde{a}_{s,k^\prime}(t_2) \right\}
\right\rangle \\
+ \left\{ t_1 \leftrightarrow t_2, t^\prime \leftrightarrow t^{\prime \prime} \right\}.
\end{multline}
This formula reveals the general structure of the Keldysh Green functions for bosonic operators: all time orderings are allowed, provided the leftmost operator is classical, and each classical (quantum) operator appears in a commutator (anticommutator) with the operators to its left, i.e., the operators with larger time arguments.
This structure will become important in the perturbative calculations in the next Section.
Thus we can write
\begin{equation} \label{eqn:2nd_4pt}
  \gamma_s(\omega^\prime | \omega) =
  \frac{i \eta v L}{\omega}
  \int \frac{\text{d} \Omega}{2\pi}
  G^{ccqq}_{\tilde{a}^\dagger_{s,k^\prime}; \tilde{a}_{s,k^\prime}; \tilde{\rho}_s (x_{in}); \tilde{\rho}_s (x_{in})} (\Omega, \omega+i\eta, -\omega+ i\eta),
\end{equation}
where again frequency arguments are in correspondence with time arguments in the previous equation (the same convention will be followed for other four-point functions below).

\begin{figure}
\includegraphics[width=16cm,height=!]{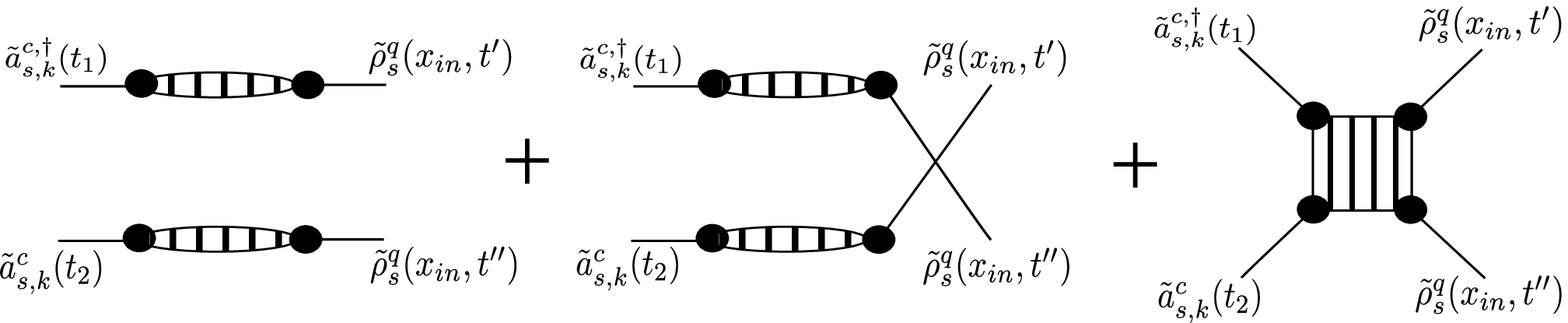}
\caption{\label{fig:diagrams}
Schematic representation of the diagrams contributing to the four-point Keldysh Green function~(\ref{eqn:gccqq}).
Lines represent decoupled-lead propagators, and
filled circles correspond to the spin-boson coupling $H_\alpha = - \pi v \alpha \tilde{\rho}_s(0) S_z$ [cf.\ Eq.~(\ref{eqn:h_sb})].
The dashed ellipses and square represent two and four point correlation functions of $S_z$, respectively, calculated to all orders in $H_\alpha$.
See the text for further details.
}
\end{figure}

Since the unitary transformations [such as $\mathcal{U}$, defined before Eq.~(\ref{eqn:h_sb}) in the main text] used to map between the different forms of the quantum impurity Hamiltonian [Eqs.~(\ref{eqn:h_lm}), (\ref{eqn:h_sb}), and (\ref{eqn:h_k}) in the main text, as well as Eq.~(\ref{eqn:h_I})] change the values of the charge densities $\rho_\ell(x)$ or the current densities $\propto \partial_x \phi_\ell(x)$ only locally, at $x=0$, they do not affect the definitions of the scattering amplitudes.
Thus, the inelastic spectrum $\gamma_s(\omega^\prime|\omega)$ [defined by Eq.~(\ref{eqn:2nd_3pt}) or Eq.~(\ref{eqn:2nd_4pt})], as well as the elastic scattering amplitudes $r_\ell(\omega)$ and $t_\ell(\omega)$ [defined by Eq.~(\ref{eqn:t_l}) in the main text], can be calculated using any of these forms of the Hamiltonian which is more convenient.
In the rest of this Section we will employ the spin-boson Hamiltonian, Eq.~(\ref{eqn:h_sb}) in the main text.

We will now show how the four-point Keldysh correlator appearing in Eq.~(\ref{eqn:2nd_4pt}) can be written in terms of local four-point spin correlation functions. This can be done using the Keldysh path integral formalism, and integrating out the lead degrees of freedom.
Alternatively, one may apply Keldysh perturbation theory to all orders in the spin-boson coupling term, $H_\alpha = - \pi v \alpha \tilde{\rho}_s(0) S_z$
\cite{Sfn:wick}.
Then, the four-point correlator defined by Eq.~(\ref{eqn:gccqq}) can be written as a sum of disconnected and connected diagrams, as depicted in Fig.~\ref{fig:diagrams}.
The former represent elastic scattering, and therefore vanish unless $\omega^\prime=\omega$. They can be shown to reproduce the square of the absolute values of the elastic scattering coefficients, Eqs.~(\ref{eqn:tmatrix})--(\ref{eqn:t_chi}) in the main text.
Since we are concerned here with inelastic scattering, we will rather concentrate only on the connected diagrams. These can be written as a product of four legs, representing two-point correlation functions of the lead operators calculated for a decoupled lead, multiplied by a four-point connected correlation function of $S_z$, calculated with the full spin-boson Hamiltonian, Eq.~(\ref{eqn:h_sb}) in the main text. We thus arrive at the following expression:
\begin{multline} \label{eqn:gccqq_long}
   G^{ccqq}_{\tilde{a}^\dagger_{s,k^\prime}; \tilde{a}_{s,k^\prime}; \tilde{\rho}_s (x_{in});\tilde{\rho}_s (x_{in})} (\Omega, \omega+i\eta, -\omega+ i\eta) = 
   \pi^4 v^4 \alpha^4
   G^{\text{hl}, qc}_{\tilde{\rho}_s (x_{in}); \tilde{\rho}_s (0)} (-\omega - i\eta)
   G^{\text{hl}, qc}_{\tilde{\rho}_s (x_{in}); \tilde{\rho}_s (0)} (\omega - i\eta)
   \\
   \qquad\qquad
   \times
   \left\{
   G^{\text{hl}, c q}_{\tilde{a}^\dagger_{s,k^\prime}; \tilde{\rho}_s(0)} (\Omega + \omega + i\eta)
   G^{\text{hl}, c q}_{\tilde{a}_{s,k^\prime}; \tilde{\rho}_s(0)} (-\Omega - \omega + i\eta)
   G^{c c q q}_{S_z;S_z;S_z;S_z}(\Omega, \omega + i\eta, -\omega + i\eta)
   \right. \\ \left.
   \qquad\qquad\qquad
   +
   G^{\text{hl}, c q}_{\tilde{a}^\dagger_{s,k^\prime}; \tilde{\rho}_s (0)} (\Omega + \omega + i\eta)
   G^{\text{hl}, c c}_{\tilde{a}_{s,k^\prime}; \tilde{\rho}_s (0)} (-\Omega - \omega + i\eta)
   G^{c q q q}_{S_z;S_z;S_z;S_z}(\Omega, \omega + i\eta, -\omega + i\eta)
   \right. \\ \left.
   +
   G^{\text{hl}, c c}_{\tilde{a}^\dagger_{s,k^\prime}; \tilde{\rho}_s (0)} (\Omega + \omega + i\eta)
   G^{\text{hl}, c q}_{\tilde{a}_{s,k^\prime}; \tilde{\rho}_s (0)} (-\Omega - \omega + i\eta)
   G^{q c q q}_{S_z;S_z;S_z;S_z}(\Omega, \omega + i\eta, -\omega + i\eta)
  \right\}.
\end{multline}
Since the four-point spin correlators are automatically connected for $\omega^\prime = \omega$, we do not need to specify this explicitly.

The disconnected lead correlation appearing in the above equation are given by
\begin{align}
  \label{eqn:ghl1}
  G^{\text{hl}, qc}_{\tilde{\rho} (x_{in}); \tilde{\rho}(0)} (\bar{\omega}) = & \frac{i \bar{\omega}}{\pi v^2} e^{i \bar{\omega} x_{in} / v}, \\
  \label{eqn:ghl2}
  G^{\text{hl}, c q/q c}_{\tilde{a}^\dagger_{k^\prime}; \tilde{\rho}(0)} (\bar{\omega}) = &
  -i \sqrt{\frac{k^\prime}{\pi L}}
  \frac{1}{\bar{\omega} + v k^\prime \pm i\eta}, \\
  \label{eqn:ghl3}
  G^{\text{hl}, c q/q c}_{\tilde{a}_{k^\prime}; \tilde{\rho}(0)} (\bar{\omega}) = &
  -i \sqrt{\frac{k^\prime}{\pi L}}
  \frac{1}{\bar{\omega} - v k^\prime \pm i\eta}, \\
  \label{eqn:ghl4}
  G^{\text{hl}, c c}_{\tilde{a}^\dagger_{s, k^\prime}; \tilde{\rho}_s (0)} (\bar{\omega}) = &
  \coth \frac{\bar{\omega}}{2 T}
  \left[ G^{\text{hl}, c q}_{\tilde{a}^\dagger_{s,k^\prime}; \tilde{\rho}_s (0)} (\bar{\omega}) - G^{\text{hl}, q c}_{\tilde{a}^\dagger_{s,k^\prime}; \tilde{\rho}_s (0)} (\bar{\omega}) \right],
\end{align}
where the last equation stems from the fluctuation-dissipation theorem. A similar relation holds for
$G^{\text{hl},cc}_{\tilde{a}_{s,k^\prime}; \tilde{\rho}_s (0)} (\bar{\omega})$.

The only terms in Eq.~(\ref{eqn:gccqq_long}) that are singular, and thus survive when the limit $\eta \to 0^+$ is taken in Eq.~(\ref{eqn:2nd_4pt}) , are those which contain the product
$G^{\text{hl}, c q}_{\tilde{a}^\dagger_{s,k^\prime}; \tilde{\rho}_s(0)} (\Omega + \omega + i\eta) G^{\text{hl}, c q}_{\tilde{a}_{s,k^\prime}; \tilde{\rho}_s(0)} (-\Omega -\omega + i\eta)$.
By Eqs.~(\ref{eqn:ghl2})--(\ref{eqn:ghl3}), this product, together with the prefactor of $\eta$ from Eq.~(\ref{eqn:2nd_4pt}), gives in that limit
\begin{equation}
  \eta G^{\text{hl}, c q}_{\tilde{a}^\dagger_{s,k^\prime}; \tilde{\rho}_s(0)} (\Omega + \omega + i\eta) G^{\text{hl}, c q}_{\tilde{a}_{s,k^\prime}; \tilde{\rho}_s(0)} (-\Omega -\omega + i\eta)
  \to
  \frac{\omega^\prime}{v L} \delta(\Omega + \omega + \omega^\prime).
\end{equation}
One may then immediately perform the integral over $\Omega$ in Eq.~(\ref{eqn:2nd_4pt}).
Plugging the result into Eq.~(\ref{eqn:gamma_ells}) we are left with:
\begin{multline} \label{eqn:gamma_4pt_app}
\gamma_{\ell^\prime | \ell} (\omega^\prime | \omega) =
\frac{i \pi}{2} \alpha_\ell^2 \alpha_{\ell^\prime}^2 \omega \omega^\prime
\times \\
\left\{
G^{c c q q}_{S_z;S_z;S_z;S_z}(\omega^\prime + \omega, -\omega, \omega)
- \coth \left( \frac{\omega^\prime}{2T} \right)
\left[
G^{cqqq}_{S_z;S_z;S_z;S_z}(\omega^\prime + \omega, -\omega, \omega)
-
G^{qcqq}_{S_z;S_z;S_z;S_z}(\omega^\prime + \omega, -\omega, \omega)
\right]
\right\}.
\end{multline}
Hence, inelastic scattering involves higher order local correlators than the elastic amplitudes:
$G^{c c q q}_{S_z;S_z;S_z;S_z}$, the
second order response of $S_z$-$S_z$ correlations to the application
of a local magnetic field, as well as
$G^{cqqq}_{S_z;S_z;S_z;S_z}$, the third order local spin susceptibility.
It can thus yield more information about the quantum impurity dynamics than elastic scattering can.

The r.h.s. of Eq.~(\ref{eqn:gamma_4pt_app}) can be evaluated exactly at the Toulouse point, $\alpha=1$, where the Kondo problem [Eq.~(\ref{eqn:h_k}) in the main text] is equivalent to a noninteracting resonant level coupled to a spinless fermionic bath \cite{Sgogolin, Sweiss, Shewson},
\begin{equation} \label{eqn:h_rlm}
  H_\text{RLM} = \sum_k v k c_k^\dagger c_k
  + \varepsilon_0 \left( d^\dagger d - \tfrac{1}{2} \right)
  + t_0 d^\dagger \sum_k c_k + \text{H.c.},
\end{equation}
where $d^\dagger$ ($c^\dagger_k$) creates an electron in the resonant level (mode $k$ of the bath),
with $S_z \rightarrow d^\dagger d - 1/2$ (and thus $\varepsilon_0 = -B_z$), as well as $t_0 = I_{xy}/(2 \sqrt{2 \pi a})$. The level width is $\Gamma = t_0^2/(2 v) = I_{xy}^2 / (16 \pi a v)$.
Since this model is quadratic, correlation functions of $S_z$ are easily calculated, using Wick's theorem and the results
\begin{align}
  G_{d;d^\dagger}^{R/A}(\bar{\omega}) & = \frac{1}{\bar{\omega} - \epsilon_0 \pm i \Gamma}, \\
  G_{d;d^\dagger}^{K}(\bar{\omega}) & = \tanh \left( \frac{\bar{\omega}}{2T} \right) \left[ G_{d;d^\dagger}^{R}(\bar{\omega}) - G_{d;d^\dagger}^{A}(\bar{\omega}) \right],
\end{align}
for the retarded, advanced, and Keldysh fermionic level Green functions, respectively \cite{Skamenev}.


The dynamic spin susceptibility is then given by
\begin{align}
\label{eqn:chi_toulouse}
  \chi_{zz}(\omega) & =
  \int_{-\infty}^{\infty} \frac{\text{d} \Omega}{2\pi}
  \frac{\Gamma}{\left[(\Omega+\omega-\varepsilon_0)^2+\Gamma^2\right]
  \left[(\Omega-\varepsilon_0)^2+\Gamma^2\right]}
  \left[
  \left(\Omega + \omega - \varepsilon_0 - i\Gamma\right)
  \tanh \left(\frac{\Omega+\omega}{2 T}\right)
  -\left(\Omega - \varepsilon_0 + i\Gamma\right)
  \tanh \left(\frac{\Omega}{2 T}\right)
  \right]
  \nonumber \\
  & =
  \frac{1}{\pi} \frac{\Gamma}{\omega (\omega + 2 i \Gamma)}
  \left[
  \psi
  \left( \frac{1}{2} + \frac{\varepsilon_0 + i\Gamma}{2 \pi i T} \right)
  + \psi
  \left( \frac{1}{2} + \frac{- \varepsilon_0 + i\Gamma}{2 \pi i T} \right)
  - \psi
  \left( \frac{1}{2} + \frac{\omega + \varepsilon_0 + i\Gamma}{2 \pi i T} \right)
  - \psi
  \left( \frac{1}{2} + \frac{\omega - \varepsilon_0 + i\Gamma}{2 \pi i T} \right)
  \right],
\end{align}
where $\psi(z)$ is the digamma function \cite{Sabramowitz}.
At zero temperature and magnetic field we get for the static susceptibility, $\chi_{zz}(0) = 1/(\pi \Gamma)$. Thus, $T_K = \pi \Gamma$ according to our definition [cf.\ the discussion before Eq.~(\ref{eqn:T_K}) of the main text].
The results of plugging Eq.~(\ref{eqn:chi_toulouse}) into Eqs.~(11) and (14) of the main text is plotted in Fig.~2 of the main text.

In addition, the correlation functions appearing in Eq.~(\ref{eqn:gamma_4pt_app}) are given by
\begin{align} \label{eqn:g_abqq_toulouse}
  G_{S_z;S_z;S_z;S_z}^{abqq} (\omega+\omega^\prime,-\omega,\omega) =
  -i\int_{-\infty}^{\infty} \frac{\text{d} \Omega}{2\pi}
  \text{Tr}
  &
  \left[
  \hat{\tau}^a \hat{G}(\omega^\prime+\Omega) \hat{\tau}^b \hat{G}(\Omega) \hat{G}(\omega+\Omega) \hat{G}(\Omega)
  \right.
  \nonumber \\
  & +
  \left.
  \hat{\tau}^a \hat{G}(\omega^\prime+\Omega) \hat{G}(\omega+\omega^\prime+\Omega) \hat{\tau}^b \hat{G}(\omega+\Omega) \hat{G}(\Omega)
  \right.
  \nonumber \\
  & +
  \left.
  \hat{\tau}^a \hat{G}(\omega^\prime+\Omega) \hat{G}(\omega+\omega^\prime+\Omega) \hat{G}(\omega^\prime+\Omega) \hat{\tau}^b \hat{G}(\Omega)
  \right.
  \nonumber \\
  & +
  \left.
  \{ \omega \leftarrow -\omega \}
  \vphantom{\hat{G}(\Omega)} \right],
\end{align}
with $a,b=c,q$, and where
\begin{equation}
  \hat{G}(\bar{\omega}) =
  \left(
  \begin{matrix}
    G_{d;d^\dagger}^{R}(\bar{\omega}) & G_{d;d^\dagger}^{K}(\bar{\omega}) \\
    0 & G_{d;d^\dagger}^{A}(\bar{\omega})
  \end{matrix}
  \right),
\end{equation}
and $\hat{\tau}^c$ is the Pauli matrix $\tau^x$, whereas $\hat{\tau}^q$ is the unit matrix. An example of the resulting inelastic spectrum is plotted in Fig.~3 of the main text.

\section{Perturbative calculation of the inelastic spectrum}
When $\omega$, $\omega^\prime$, and $|\omega-\omega^\prime|$ or $T$ are large with respect to the Kondo temperature, one may evaluate the inelastic spectrum $\gamma_{\ell^\prime | \ell} (\omega^\prime | \omega)$ perturbatively in $I_{xy} \propto E_J^L E_J^R$ [cf.\ Eq.~(\ref{eqn:bz_ejlr}) in the main text] for any value of $\alpha$, and obtain Eq.~(\ref{eqn:gamma_pert0}) in the main text. In this section we will present the details of this calculation.

In this regime it is useful to apply the transformation
$H \rightarrow \mathcal{V}^\dagger H \mathcal{V}$ with $\mathcal{V}= e^{i \alpha \tilde{\phi}_s (0) S_z}$ to the spin-boson Hamiltonian, Eq.~(\ref{eqn:h_sb}) in the main text (as argued in the previous Section, such a transformation does not affect the scattering amplitudes), so as to transfer the impurity-leads coupling into the perturbative $I_{xy}$ term,
\begin{equation}
  \label{eqn:h_I}
  H_{I} =
  \sum_{\lambda=c,s} \frac{v}{2\pi} \int_0^\infty \left\{ \left[ \partial_x \tilde{\phi}_\lambda(x) \right]^2 + \left[ \pi \tilde{\rho}_\lambda(x) \right]^2 \right\} \text{d}x
  \\
  - B_z S_z
  - \frac{I_{xy}}{4 \pi a} \left( e^{-i \alpha \tilde{\phi}_s(0)} S_+ + e^{i \alpha \tilde{\phi}_s(0)} S_- \right)
\end{equation}

Expanding the Keldysh Green functions appearing in Eq.~(\ref{eqn:gamma_4pt_app}) in $I_{xy}$,
the zeroth and first order terms vanish.
The second order terms breaks down into a products two-point boson correlator and a six-point spin correlator, to be evaluated for the Hamiltonian~(\ref{eqn:h_I}) with $I_{xy}=0$:
\begin{align}
\label{eqn:gccqq_expand}
\begin{split}
  & G^{ccqq}_{S_z;S_z;S_z;S_z}(t_1-t_2, t_1-t^\prime, t_2-t^{\prime \prime})
  = i \left( \frac{I_{xy}}{4 \pi a} \right)^2
  \times \\
  & \qquad\qquad
  \sum_{a,b=q,c}
  \int_{-\infty}^{\infty} \text{d}s_1
  \int_{-\infty}^{\infty} \text{d}s_2
  \left\langle S_z^c(t_1) S_z^c(t_2) S_z^q(t^\prime) S_z^q(t^{\prime \prime})
  S_+^a(s_1) S_-^b(s_2) \right\rangle
  \left\langle \right[e^{-i \alpha \tilde{\phi}_s(0,s_1)}\left]_{\bar{a}} \right[e^{i \alpha \tilde{\phi}_s(0,s_2)}\left]_{\bar{b}} \right\rangle,
\end{split}
\\
\label{eqn:gcqqq_expand}
\begin{split}
  & G^{cqqq}_{S_z;S_z;S_z;S_z}(t_1-t_2, t_1-t^\prime, t_2-t^{\prime \prime})
  = i \left( \frac{I_{xy}}{4 \pi a} \right)^2
  \times \\
  & \qquad\qquad
  \sum_{a,b=q,c}
  \int_{-\infty}^{\infty} \text{d}s_1
  \int_{-\infty}^{\infty} \text{d}s_2
  \left\langle S_z^c(t_1) S_z^q(t_2) S_z^q(t^\prime) S_z^q(t^{\prime \prime})
  S_+^a(s_1) S_-^b(s_2) \right\rangle
  \left\langle \right[e^{-i \alpha \tilde{\phi}_s(0,s_1)}\left]_{\bar{a}} \right[e^{i \alpha \tilde{\phi}_s(0,s_2)}\left]_{\bar{b}} \right\rangle,
\end{split}
\end{align}
whereas $G^{qcqq}_{S_z;S_z;S_z;S_z}(t_1-t_2, t_1-t^\prime, t_2-t^{\prime \prime})$ is obtained from
$G^{cqqq}_{S_z;S_z;S_z;S_z}(t_2-t_1, t_2-t^\prime, t_1-t^{\prime \prime})$
by interchanging $t_1$ and $t_2$.
Here $\bar{a}=q,c$ for $a=c,q$, respectively, and similarly for $\bar{b}$.
Therefore, the term with $a=b=c$ contains a $q$-$q$ lead correlator, and thus vanishes.

For $\omega,\omega^\prime,|\omega-\omega^\prime| \gg B_z$, one may neglect the effects of the magnetic field.
Then, the spin operators appearing in Eqs.~(\ref{eqn:gccqq_expand})--(\ref{eqn:gcqqq_expand}) are time independent.
Following the rules of the Keldysh formalism \cite{Schou85}, the corresponding spin correlators can be written as combinations of commutators and anticommutators of the spin operators, depending on the ordering of the time arguments. Most of these turn out to be zero.
The spin correlator on the r.h.s. of the Eq.~(\ref{eqn:gccqq_expand}) does not vanish only if $a=b=q$, in which case it gives
\begin{equation} \label{eqn:6s1}
  \left\langle S_z^c(t_1) S_z^c(t_2) S_z^q(t^\prime) S_z^q(t^{\prime \prime})
  S_+^q(s_1) S_-^q(s_2) \right\rangle =
  \theta(t_1-s_1) \theta(s_1-t^\prime) \theta(s_1 - t^{\prime \prime}) \theta(t^\prime - s_2) \theta(t^{\prime \prime} - s_2) \theta(s_2-t_2) + \left\{ s_1 \leftrightarrow s_2 \right\},
\end{equation}
whereas the spin correlator on the r.h.s. of the Eq.~(\ref{eqn:gcqqq_expand}) does not vanish only if $a=c$, $b=q$, when
\begin{align}
  \label{eqn:6s2}
  \left\langle S_z^c(t_1) S_z^q(t_2) S_z^q(t^\prime) S_z^q(t^{\prime \prime})
  S_+^c(s_1) S_-^q(s_2) \right\rangle =
  &
  \theta(t_1-s_2) \theta(s_2-t_2) \theta(s_2 - t^\prime) \theta(s_2-t^{\prime \prime}) \theta(t_2-s_1) \theta(t^\prime-s_1) \theta(t^{\prime \prime}-s_1) +
  \nonumber \\ &
  \theta(s_1-t_2) \theta(s_1-t^\prime) \theta(s_1-t^{\prime \prime}) \theta(t_2-s_2) \theta(t^\prime-s_2) \theta(t^{\prime \prime}-s_2) \theta(s_2-t_1),
\end{align}
or if $a=q$, $b=c$, in which case one should simply interchange $s_1$ and $s_2$ in the last equation.

Plugging Eqs.~(\ref{eqn:6s1})--(\ref{eqn:6s2}) back into Eqs.~(\ref{eqn:gccqq_expand})--(\ref{eqn:gcqqq_expand}), one can perform the integrals over $s_1$ and $s_2$, and then calculate the Fourier-transform of the results.
Using in addition the fluctuation-dissipation theorem to express all the different Keldysh lead correlators in terms of the retarded one, $\tilde{\chi}^\text{hl}_{+-} (\omega) = \dleft\langle e^{i \alpha \tilde{\phi}_s(0)}; e^{-i \alpha \tilde{\phi}_s(0)} \dright\rangle^\text{hl}_{\omega}$, we find
\begin{align}
\begin{split}
  G^{ccqq}_{S_z;S_z;S_z;S_z}(\omega+\omega^\prime, -\omega, \omega)
  = &
  \frac{4 i}{\omega^2 \omega^{\prime 2}}
  \left( \frac{I_{xy}}{4 \pi a} \right)^2 \left\{
  2\coth \left( \frac{\omega^\prime}{2T} \right)
  \text{Im} \left[ \tilde{\chi}^\text{hl}_{+-}(\omega^\prime) \right]
  -\coth \left( \frac{\omega+\omega^\prime}{2T} \right)
  \text{Im} \left[ \tilde{\chi}^\text{hl}_{+-}(\omega+\omega^\prime) \right]
  \right. \\ & \qquad \qquad \qquad \qquad \qquad \qquad \qquad \qquad \qquad \qquad \left.
  -\coth \left( \frac{\omega-\omega^\prime}{2T} \right)
  \text{Im} \left[ \tilde{\chi}^\text{hl}_{+-}(\omega-\omega^\prime) \right]
  \right\},
\end{split}
\\
\begin{split}
  G^{cqqq}_{S_z;S_z;S_z;S_z}(\omega+\omega^\prime, -\omega, \omega)
  = &
  - G^{qcqq}_{S_z;S_z;S_z;S_z}(\omega+\omega^\prime, -\omega, \omega)
  =
  \\ & \qquad
  \frac{2 i}{\omega^2 \omega^{\prime 2}}
  \left( \frac{I_{xy}}{4 \pi a} \right)^2 \Bigl\{
  2 \text{Im} \left[ \tilde{\chi}^\text{hl}_{+-}(\omega^\prime) \right]
  - \text{Im} \left[ \tilde{\chi}^\text{hl}_{+-}(\omega+\omega^\prime) \right]
  + \text{Im} \left[ \tilde{\chi}^\text{hl}_{+-}(\omega-\omega^\prime) \right]
  \Bigr\}.
\end{split}
\end{align}
Substituting these expressions into Eq.~(\ref{eqn:gamma_4pt_app}) we arrive at Eq.~(\ref{eqn:gamma_pert0}) in the main text:
\begin{align}
  \gamma_{\ell^\prime | \ell} (\omega^\prime | \omega) =
  \frac{4\pi \alpha_\ell^2 \alpha_{\ell^\prime}^2}{\omega \omega^\prime}
  \left(\frac{I_{xy}}{4 \pi a}\right)^2
  &
  \left[
  \text{Im}
  \left[ \tilde{\chi}^\text{hl}_{+-} (\omega-\omega^\prime) \right]
  \left(
  \theta(\omega-\omega^\prime)
  \Bigl\{
  \left[ 1+n_B (\omega^\prime) \right]
  \left[ 1+n_B (\omega-\omega^\prime) \right]
  - n_B (\omega^\prime) n_B (\omega-\omega^\prime)
  \Bigr\}
  \right. \right. \nonumber \\ & \qquad\qquad\qquad\qquad
  \left. \left.
  + \theta(\omega^\prime-\omega)
  \Bigl\{
  n_B (\omega^\prime) \left[ 1+n_B (\omega^\prime-\omega) \right]
  - \left[ 1+n_B (\omega^\prime)\right] n_B (\omega^\prime-\omega)
  \Bigr\}
  \right)
  \right. \nonumber \\ & \left.
  + \text{Im}
  \left[ \tilde{\chi}^\text{hl}_{+-} (\omega+\omega^\prime) \right]
  \Bigl\{
  n_B (\omega+\omega^\prime) \left[ 1+n_B (\omega^\prime) \right]
  - \left[ 1+n_B (\omega+\omega^\prime)\right] n_B (\omega^\prime)
  \Bigr\}
  \right], \nonumber
\end{align}
where \cite{Sgogolin}
\begin{equation}
  \tilde{\chi}^\text{hl}_{+-} (\Omega) =
  \frac{1}{2} \sin \left( \frac{\pi \alpha^2}{2} \right)
  \frac{1}{\omega_0} \left(\frac{2\pi T}{\omega_0}\right)^{\alpha^2-1}
  B \left( \frac{\alpha^2}{2} - i\frac{\Omega}{2\pi T}, 1-\alpha^2 \right),
\end{equation}
with $B(x,y)$ the beta function \cite{Sabramowitz}.
Thus, $\text{Im}[\tilde{\chi}^\text{hl}_{+-} (\Omega)] \propto [\max(\Omega,T))]^{\alpha^2-1}$ for small $\Omega$ and $T$.
As mentioned in the main text, the first line of Eq.~(\ref{eqn:gamma_pert0}) describes a process where an incoming photon at frequency $\omega$ is absorbed by the quantum impurity, and a photon at frequency $\omega^\prime<\omega$, plus additional photons whose energies sum up to $\omega-\omega^\prime$ are emitted (this is the only process allowed at zero temperature) and the reverse process.
Similarly, the second line describes a process where a photon at frequency $\omega^\prime>\omega$ is absorbed, and a photon at frequency $\omega$, as well as photons whose frequencies sum up to $\omega^\prime-\omega$ are emitted and vice versa.
Finally, the third line describes a process where photons whose frequencies sum up to $\omega^\prime+\omega$ are absorbed, and photons at frequencies $\omega$ and $\omega^\prime$ are emitted and vice versa.

The structure of Eq.~(\ref{eqn:gamma_pert0}) in the main text thus suggests that it can be obtained from a kinetic equation.
Indeed, one can write down the Boltzmann equation for the average mode occupations $\tilde{n}_q \equiv \langle \tilde{a}^\dagger_{s,q} \tilde{a}_{s,q}\rangle$, accounting for all the possible multiphoton scattering processes to second order in $I_{xy}$. The corresponding probabilities can be obtained by Fermi's golden rule from the Hamiltonian~(\ref{eqn:h_I}), after expanding the exponents in the last term of the Hamiltonian to all orders in the bosonic fields
\cite{Sfn:bib_nphoton}:
\begin{multline} \label{eqn:boltzmann}
\frac{\text{d} \tilde{n}_q}{\text{d} t} =
2\pi \left( \frac{I_{xy}}{4 \pi a} \right)^2 \frac{\pi}{q L}
\sum_{N,N^\prime=1}^{\infty}
\frac{\alpha^{2(N+N^\prime + 1)}}{N! N^\prime !}
\times \\
\left[
(1+\tilde{n}_q)
\int \frac{\text{d} q_1}{q_1} \cdots \int \frac{\text{d} q_N}{q_N}
\int \frac{\text{d} q^\prime_1}{q^\prime_1} \cdots \int \frac{\text{d} q^\prime_{N^\prime}}{q^\prime_{N^\prime}}
\tilde{n}_{q_1} \cdots \tilde{n}_{q_N} (1+\tilde{n}_{q^\prime_1}) \cdots (1+\tilde{n}_{q^\prime_{N^\prime}})
\delta (\omega_{q_1} + \cdots + \omega_{q_N} - \omega_{q^\prime_1} - \cdots -\omega_{q^\prime_{N^\prime}}- \omega_q)
\right. \\ \left.
- \tilde{n}_q
\int \frac{\text{d} q_1}{q_1} \cdots \int \frac{\text{d} q_N}{q_N}
\int \frac{\text{d} q^\prime_1}{q^\prime_1} \cdots \int \frac{\text{d} q^\prime_{N^\prime}}{q^\prime_{N^\prime}}
\tilde{n}_{q_1} \cdots \tilde{n}_{q_N} (1+\tilde{n}_{q^\prime_1}) \cdots (1+\tilde{n}_{q^\prime_{N^\prime}})
\delta (\omega_q + \omega_{q_1} + \cdots + \omega_{q_N} - \omega_{q^\prime_1} - \cdots -\omega_{q^\prime_{N^\prime}})
\right],
\end{multline}
where $\omega_Q \equiv v Q$.
In equilibrium [in the absence of the external driving $V(t)$], the mode occupations are given by the Bose-Einstein distribution, $\tilde{n}_Q = n_B(\omega_Q)$.
In order to find the rate of change of occupation of mode
$q = k^\prime$ by the ac excitation $V(t)$ to second order in $I_{xy}$, one should substitute on the right hand side of Eq.~(\ref{eqn:boltzmann}) the equilibrium (Bose-Einstein) occupations for all the modes, except for the mode with wavevector $k = \omega/v$, whose occupation is modified by $n_k^V$ by the ac source $V(t)$; thus, $n_Q = n_B(\omega_Q) + (\pi/L) n_k^V \delta(Q-k)$.
Multiplying the resulting rate by the photon density of states $L/(\pi v)$, and dividing by 
the incoming flux of photons of frequency $\omega$ [i.e., $n_k^V v / (2 L)$],
we recover our previous result, Eq.~(\ref{eqn:gamma_pert0}) in the main text, if we employ the following relations:
\begin{multline}
  \left[ 1+n_B (\Omega) \right]
  \text{Im} \dleft\langle e^{i \alpha \phi(0)}; e^{-i \alpha \phi(0)} \dright\rangle_{\Omega}^\text{hl} =
  \pi
  \sum_{N,N^\prime=1}^{\infty} \frac{\alpha^{N+N^\prime}}{N! N^\prime !}
  \int \frac{\text{d} q_1}{q_1} \cdots \int \frac{\text{d} q_N}{q_N}
  \int \frac{\text{d} q^\prime_1}{q^\prime_1} \cdots \int \frac{\text{d} q^\prime_{N^\prime}}{q^\prime_{N^\prime}}
  \times \\
  n_B(\omega_{q_1}) \cdots n_B(\omega_{q_N}) \left[1+n_B(\omega_{q^\prime_1})\right] \cdots \left[1+n_B(\omega_{q^\prime_{N^\prime}})\right]
  \delta (\Omega + \omega_{q_1} + \cdots + \omega_{q_N} - \omega^\prime
  - \omega_{q^\prime_1} \cdots -\omega_{q^\prime_{N^\prime}}),
\end{multline}
\begin{multline}
  n_B (\Omega)
  \text{Im} \dleft\langle e^{i \alpha \phi(0)}; e^{-i \alpha \phi(0)} \dright\rangle_{\Omega}^\text{hl} =
  \pi
  \sum_{N,N^\prime=1}^{\infty} \frac{\alpha^{N+N^\prime}}{N! N^\prime !}
  \int \frac{\text{d} q_1}{q_1} \cdots \int \frac{\text{d} q_N}{q_N}
  \int \frac{\text{d} q^\prime_1}{q^\prime_1} \cdots \int \frac{\text{d} q^\prime_{N^\prime}}{q^\prime_{N^\prime}}
  \times \\
  n_B(\omega_{q_1}) \cdots n_B(\omega_{q_N}) \left[1+n_B(\omega_{q^\prime_1})\right] \cdots \left[1+n_B(\omega_{q^\prime_{N^\prime}})\right]
  \delta (\omega_{q_1} + \cdots + \omega_{q_N} - \omega^\prime
  - \omega_{q^\prime_1} \cdots -\omega_{q^\prime_{N^\prime}}- \Omega).
\end{multline}

\section{Inelastic scattering in the small $\alpha$ limit}
In this Section we will analyze inelastic photon scattering in the limit of small $\alpha$ at zero temperature.
In that regime it is useful to use the spin-boson version of the Hamiltonian, Eq.~(\ref{eqn:h_sb}) in the main text.
We will start from the case $B_z=0$.
Then we have a two-level system ($S_x = \pm 1/2$), where the two levels are separated by $T_K = E_J^{LR}$, and weakly coupled to the bath of photons.
Since every photon emission or absorption flips the impurity spin, the inelastic process which is lowest-order in $\alpha$ and leaves the two-level system in its ground state involves four photons.
The amplitude for a photon at frequency $\omega$ incoming in lead $\ell$ to scatter into photons of frequencies $\omega^\prime$, $\omega_1$, and $\omega_2 = \omega - \omega^\prime - \omega_1$ outgoing into leads $\ell^\prime$, $\ell_1$ and $\ell_2$, respectively, is, to the lowest order in $\alpha$, a sum over the partial amplitudes of the $4!=24$ different orderings of the absorption of the single incoming photon and the emission of the three outgoing ones. Squaring this total amplitude and multiplying by the appropriate density of states factors we find the cross section
\begin{multline} \label{eqn:scattering_inelastic}
  \gamma_{\ell^\prime, \ell_1, \ell_2 | \ell} (\omega^\prime, \omega_1, \omega_2 | \omega) =
  \frac{\pi^2}{2}
  \alpha_{\ell}^2 \alpha_{\ell^\prime}^2 \alpha_{\ell_1}^2 \alpha_{\ell_2}^2
  \left| E_J^{LR} \right|^2
  \omega \omega^\prime \omega_1 \omega_2
  \times \\
  \left|
  \frac{\omega \omega^\prime \omega_1 \omega_2 - \left(\tilde{E}_J^{LR}\right)^2 \left( \omega^{\prime 2} + \omega_1^2 + \omega_2^2 + \omega^\prime \omega_1 + \omega^\prime \omega_2 + \omega_1 \omega_2 \right) + 3\left(\tilde{E}_J^{LR}\right)^4 }
  {(\omega-\tilde{E}_J^{LR}) (\omega+\tilde{E}_J^{LR}) (\omega^\prime-\tilde{E}_J^{LR}) (\omega^\prime+\tilde{E}_J^{LR}) (\omega_1-\tilde{E}_J^{LR}) (\omega_1+\tilde{E}_J^{LR}) (\omega_2-\tilde{E}_J^{LR}) (\omega_2+\tilde{E}_J^{LR})}
  \right|^2,
\end{multline}
where $\tilde{E}_J^{LR} = E_J^{LR}[1-(\alpha^2/2) \ln(\omega/E_J^{LR})] + i \Gamma_J^{LR}$ accounts for the shift and finite lifetime (broadening) of the excited impurity state, with $\Gamma_J^{LR} = \pi \alpha^2 E_J^{LR}/4$ (The shift in the real part of $E_J^{LR}$ corresponds to the change in $T_K$, Eq.~(\ref{eqn:T_K}) in the main text, calculated to order $\alpha^2$).
Thus, while the cross section is only of order $\alpha^8$ for small $\alpha$ (for fixed $\alpha_L/\alpha_R$), it displays peaks of height $\propto \alpha^4$ and width $\propto \alpha^2$ whenever one of the frequencies is close to $E_J^{LR}$.
It should be noted that having more than one of the outgoing frequencies $\omega^\prime$, $\omega_1$, and $\omega_2$ close to $E_J^{LR}$ does not lead to even higher peaks, since the numerator in Eq.~(\ref{eqn:scattering_inelastic}) vanishes in that case.

Integrating over $\omega_1$ and summing over $\ell_{1,2}$ we find the four-photon process contribution to the inelastic spectrum
\begin{equation} \label{eqn:scattering_inelastic_int}
  \gamma^{(4)}_{\ell^\prime | \ell} (\omega^\prime | \omega) =
  \sum_{\ell_1,\ell_2=L,R} \int_0^{\omega-\omega^\prime}
  \gamma_{\ell^\prime, \ell_1, \ell_2 | \ell} (\omega^\prime, \omega_1, \omega-\omega^\prime-\omega_1 | \omega) \text{d}\omega_1.
\end{equation}
Let us discuss the main features in the dependence of $\gamma^{(4)}_{\ell^\prime | \ell} (\omega^\prime | \omega)$ on $\omega$, $\omega^\prime$ and $\alpha$.
When all frequencies are small with respect to $E_J^{LR}$, no resonance contributes, leading to the second term in Eq.~(\ref{eqn:gamma_small_omega}) in the main text, with $a_\omega(\alpha) = 3 \pi^2 \alpha^4/4$.
In that case, therefore, the spectrum $\gamma_{\ell^\prime | \ell} (\omega^\prime | \omega)$, as well as the total inelastic scattering probability $\gamma_{\ell} (\omega)$ [obtained through the sum rule, Eq.~(\ref{eqn:sum_rule}) in the main text], are very small, of order $\alpha^8$.

\begin{figure}
\includegraphics[width=12cm,height=!]{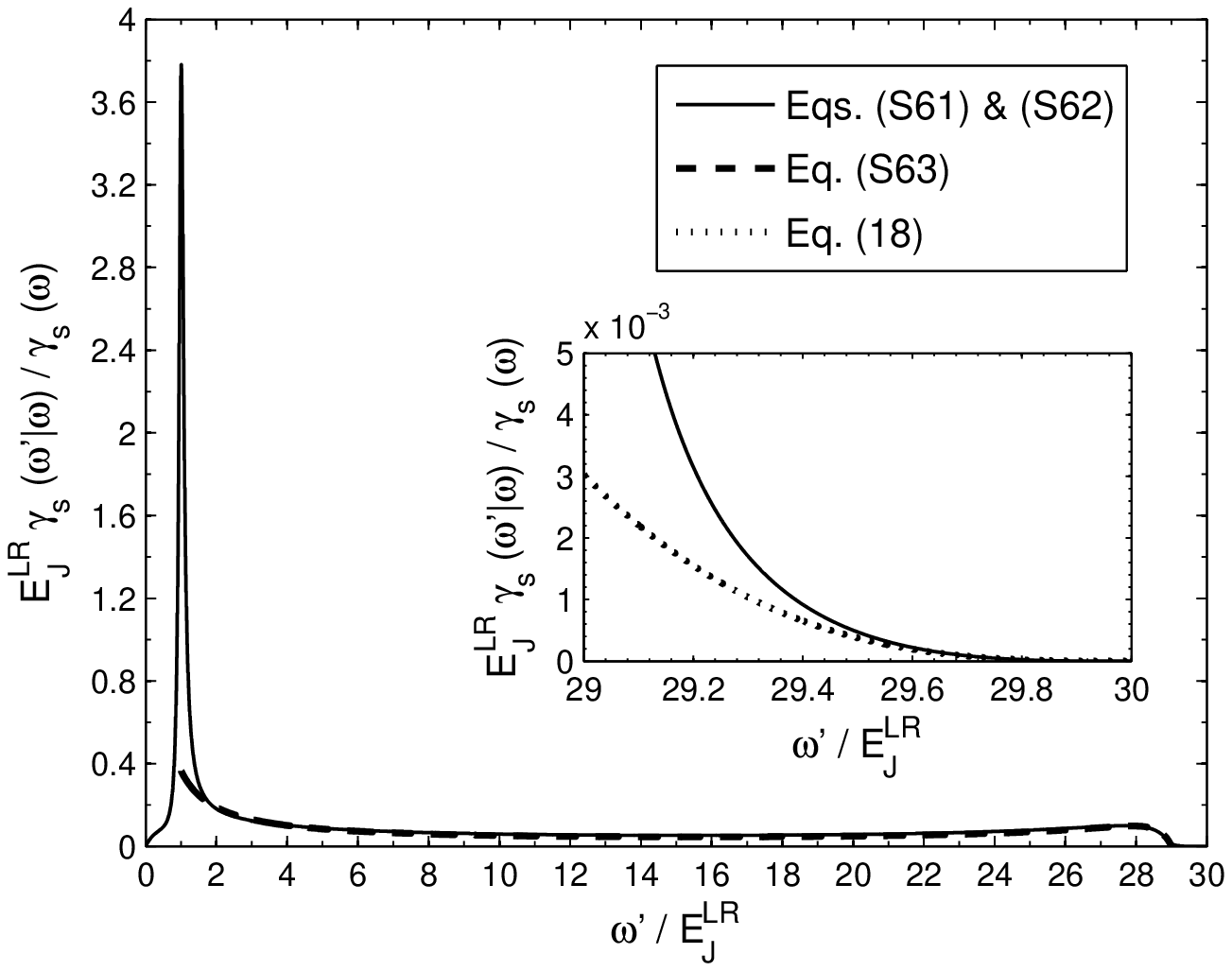}
\caption{\label{fig:spectrum_small_alpha_numerical}
$\gamma_s(\omega^\prime|\omega) \equiv \sum_{\ell,\ell^\prime=L,R} \gamma_{\ell^\prime|\ell} (\omega^\prime|\omega)$, representing the inelastic spectrum summed over the incoming and outgoing leads, normalized by the total probability summed over the incoming lead,
$\gamma_s(\omega) \equiv \sum_{\ell=L,R} \gamma_\ell(\omega)$,
for $\alpha^2=0.1$, $\omega/E_J^{LR}=30.0$, and $B_z=T=0$.
In this regime
$\gamma_s(\omega) \sim \alpha^6 (E_J^{LR}/\omega)^2 \ln (\omega/E_J^{LR})$.
The inset is a zoom-in into the region $\omega-\omega^\prime < E_J^{LR}$.
The continuous line is the exact result (to leading order in $\alpha$), obtained from numerical evaluation of Eq.~(\ref{eqn:scattering_inelastic_int}) together with Eq.~(\ref{eqn:scattering_inelastic}). The dashed line corresponds to Eq.~(\ref{eqn:scattering_inelastic_pert}), valid for $E_J^{LR} \ll \omega^\prime < \omega - E_J^{LR}$, and the dotted line corresponds to Eq.~(\ref{eqn:gamma_small_omega_omegap}) in the main text, valid for $\omega-\omega^\prime \ll E_J^{LR}$.
Note the nonmonotonic behavior for $\omega^\prime>T_K$, as compared with Fig.~3 of the main text. This nonmonotonicity, and the resulting broad peak around $\omega-\omega^\prime \sim T_K$, are expected for any $\alpha<1$ by Eq.~(\ref{eqn:gamma_pert}) in the main text.
See the text for further details.
}
\end{figure}

For $\omega \gg E_J^{LR}$ (more precisely, $\omega > 2 E_J^{LR}$), the behavior is richer, as depicted in Fig.~\ref{fig:spectrum_small_alpha_numerical}.
For $\omega^\prime > \omega - E_J^{LR}$ none of the frequencies is close to a pole, and the spectrum is still $\propto \alpha^8$, corresponding to the second term in Eq.~(\ref{eqn:gamma_small_omega_omegap}) in the main text,
with
$a_\omega^\prime(\alpha) = \pi^2 \alpha^4 /12$.
For $\omega^\prime < \omega - E_J^{LR}$ the integration over $\omega_1$ includes the regions $\omega_{1,2} \approx E_J^{LR}$, so the spectrum is $\propto \alpha^6$ in most of this range, except for a peak of height $\propto \alpha^4$ and width $\propto \alpha^2$ when $\omega^\prime \approx E_J^{LR}$.
Away from that peak, in the regime $E_J^{LR} \ll \omega^\prime < \omega - E_J^{LR}$ the calculation can be carried out explicitly to the lowest order in $\alpha$, leading to,
\begin{equation}
  \label{eqn:scattering_inelastic_pert}
  \gamma^{(4)}_{\ell^\prime | \ell} (\omega^\prime | \omega) =
  \pi^2
  \alpha^2 \alpha_{\ell}^2 \alpha_{\ell^\prime}^2
  \frac{\left( E_J^{LR} \right)^2 \left( \omega - \omega^\prime - E_J^{LR} \right)}
  {\omega \omega^{\prime} \left( \omega - \omega^\prime \right)^2}.
\end{equation}
It should be noted that Eq.~(\ref{eqn:scattering_inelastic_pert}) agrees with Eq.~(\ref{eqn:gamma_pert}) in the main text in their common domain of applicability, i.e., lowest order in $\alpha$ and the range $E_J^{LR} \ll \omega^\prime \ll \omega - E_J^{LR}$.
By Eq.~(\ref{eqn:sum_rule}) in the main text, this latter range gives the dominant contribution to the total inelastic scattering probability $\gamma_\ell(\omega)$ for $\omega \gg E_J^{LR}$. Eq.~(\ref{eqn:scattering_inelastic_pert}) results in $\gamma_\ell(\omega) \sim (E_J^{LR}/\omega)^2 \alpha^6 \ln (\omega/E_J^{LR})$ for $\alpha^2 \ln (\omega/E_J^{LR}) \ll 1$, whereas Eq.~(\ref{eqn:gamma_pert}) in the main text shows that $\gamma_\ell(\omega) \sim \alpha^4 (E_J^{LR}/\omega)^2$ for $\alpha^2 \ln (\omega/E_J^{LR}) \gg 1$.

In the regime $E_J^{LR} <\omega < 2E_J^{LR}$ a similar analysis leads to a total inelastic probability $\propto \alpha^6$.
Finally, when $\omega$ itself is resonant, $\omega \approx E_J^{LR}$, both $\gamma_{\ell^\prime | \ell} (\omega^\prime | \omega)$ and $\gamma_\ell(\omega)$ are $\propto \alpha^4$.
The peaks when one of the frequencies $\omega^\prime$, $\omega_1$, $\omega_2$, is also close to $E_J^{LR}$ are suppressed here by the frequency factors in the first line of Eq.~(\ref{eqn:scattering_inelastic}), since the other two frequencies must be close to zero in this case.

Turning on a finite magnetic field $B_z$, three-photon processes become possible. To lowest order in $B_z/E_J^{LR}$ their contribution to the inelastic spectrum is
\begin{multline} \label{eqn:scattering_inelastic_bz}
  \gamma^{(3)}_{\ell^\prime | \ell} (\omega^\prime | \omega) =
  \frac{\pi^2}{2}
  \alpha_{\ell}^2 \alpha_{\ell^\prime}^2 \alpha^2
  B_z^2
  \left|\tilde{E}_J^{LR}\right|^2
  \omega \omega^\prime (\omega - \omega^\prime)
  \times \\
  \left|
  \frac{\omega^2 + \omega^{\prime 2} - \omega \omega^\prime - 3\left(\tilde{E}_J^{LR}\right)^2 }
  {(\omega-\tilde{E}_J^{LR}) (\omega+\tilde{E}_J^{LR}) (\omega^\prime-\tilde{E}_J^{LR}) (\omega^\prime+\tilde{E}_J^{LR}) (\omega-\omega^\prime-\tilde{E}_J^{LR}) (\omega-\omega^\prime+\tilde{E}_J^{LR}) }
  \right|^2.
\end{multline}
At small frequencies we now recover the first term in Eq.~(\ref{eqn:gamma_small_omega}) in the main text,
with $a_B(\alpha) = 9 \pi^2 \alpha^2/2$,
i.e., $\gamma_{\ell^\prime|\ell}(\omega^\prime|\omega) \propto \alpha^6$.
For $\omega>E_J^{LR}$ the spectrum has two peaks, at $\omega^\prime \approx E_J^{LR}$ and $\omega-\omega^\prime \approx E_J^{LR}$, both of height and width $\propto \alpha^2$, leading to total inelastic probability $\propto \alpha^4$, whereas for $\omega-\omega^\prime \ll E_J^{LR}$ we recover the first term in Eq.~(\ref{eqn:gamma_small_omega_omegap}) in the main text for
$\omega \gg E_J^{LR}$,
with $a_B^\prime(\alpha) = \pi^2 \alpha^2/2$.
Finally, for $\omega \approx E_J^{LR}$ we have $\gamma_{\ell^\prime | \ell} (\omega^\prime | \omega) \propto \alpha^2$, with narrow peaks at $\omega^\prime \sim \Gamma_J^{LR}$ and $\omega - \omega^\prime \sim \Gamma_J^{LR}$, resulting in $\gamma_\ell(\omega) \propto \alpha^2 \ln (1/\alpha^2)$. On the other hand, all these values are suppressed by a factor $\sim (B_z/E_J^{LR})^2$.

\newpage
\end{widetext}


\begin{thebibliography}{99}
\bibitem{fn:ejlr_renormalization}
The spin flip amplitude $E_J^{LR}$ should actually be multiplied by a factor of $e^{-\alpha^2/2}$ [with $\alpha$ defined in Eq.~(\ref{eqn:alpha})] which is close to unity, to account for its renormalization \cite{weiss} by the integration out of the bosonic modes with frequencies between $\sim \omega_0$ and $\sim v/a$, whose dispersion is nonlinear.
\bibitem{fn:alpha_lr}
We maintain the distinction between $\alpha_L$ and $\alpha_R$ in the following since they may differ if the hierarchy of capacitances is different.
\bibitem{fn:lead_length}
As discussed above, the lead lengths need also be larger than $a\sqrt{C/C_g}$ but small enough to make phase slips negligible.
\bibitem{fn:chi_toulouse}
At $\alpha=1$ (the standard Toulouse point \cite{gogolin,weiss,hewson}) Eq.~(\ref{eqn:chi_high_omega}) is modified by a logarithmic factor (See SM, Sec.~SM.D \cite{sm}).
\bibitem{fn:boundary_condition_nozieres_basis}
Remember that $\partial_x \tilde{\phi}_s (0)$ is zero, due to the absence of net current through the impurity.
In addition, it should be noted that the bosonic Nozi\`{e}res description involves bosonic fields corresponding to fermionic states modified by the Kondo fixed point phase shift.
\end{thebibliography}
\begin{thebibliography}{99}
\bibitem[S\theSbib]{Sfn:scatter_phi}
One may use correlators involving the fields $\phi_\ell(x)$ to obtain the same results.

\stepcounter{Sbib}
\bibitem[S\theSbib]{Sfn:V_rhos}
Because of this normalization we are allowed to choose to couple only the spin density field $\rho_s(x_{in})$ to $V(t)$.

\stepcounter{Sbib}
\bibitem[S\theSbib]{Sfn:bib_nphoton}
The number of photons involved in such a scattering process can be odd even in the absence of a magnetic field. This may seem at first sight to contradict the statements made in the main text in the discussion of the small-frequency limit. However, one should note that the $I_{xy}$ term of the Hamiltonian~(\ref{eqn:h_I}) changes the state of the impurity. Restoring the impurity state involves the emission of additional photons, and would bring their total number to an even value, as explained in the main text. This may be ignored in the current perturbative evaluation of the inelastic spectrum $\gamma_{\ell^\prime|\ell}(\omega^\prime|\omega)$. 
\end{thebibliography}
\end{document}